\documentclass[iop]{emulateapj}
\usepackage{amsmath} 
\usepackage[utf8x]{inputenc}
\usepackage{natbib}

\usepackage{hyperref}
\usepackage{color}
\definecolor{MidnightBlue}	{RGB}{ 25,  25, 112}
\definecolor{Navy}		{RGB}{  0,   0, 128}
\hypersetup{colorlinks=true, linkcolor=Navy, citecolor=Navy, urlcolor=Navy}

\providecommand{\e}[1]{\ensuremath{\times10^{#1}}}

\def \yr        {{\rm\ yr}}

\def \kev       {{\rm\ keV}}

\def \logTd6 {\hbox{log$( T/6 \kev)$} }

\def \pMpc	{\,\mathrm{pMpc}}
\def \cMpc	{\,\mathrm{cMpc}}

\def \ckpc	{\,\mathrm{ckpc}}
	
\def \Myr	{\,\mathrm{Myr}}
\def \Gyr	{\,\mathrm{Gyr}}
\def \Msun	{\,\mathrm{M_\odot}}

\def \GqsoH	{\Gamma^\mathrm{HI}_\mathrm{QSO}}
\def \GqsoHe	{\Gamma^\mathrm{HeII}_\mathrm{QSO}}

\def \GuvbH	{\Gamma^\mathrm{HI}_\mathrm{UVB}}
\def \GuvbHe	{\Gamma^\mathrm{HeII}_\mathrm{UVB}}

\def \tQ	{t_\mathrm{Q}}
\def \tTO	{t_\mathrm{age}}
\def \xiTPE	{\xi}
\def \Oobsc	{\Omega_\mathrm{obsc}}
\def \fobsc	{f_\mathrm{obsc}}
\def \F		{F}				
\def \CT	{C}			
\def \CML	{X}			
\def \SE	{\mathcal{S}}			
\def \EX	{\mathcal{T}}			
\def \CO	{\mathcal{C}}			
\def \BG	{\mathcal{B}}			

\newcommand{\lya}{{\rm Ly}\alpha}

\newcommand{\Nyx}{\textsc{Nyx} }

\DeclareMathAlphabet{\mathsc}{OT1}{cmr}{m}{sc}
\DeclareRobustCommand{\ion}[2]{%
  \relax
  \ifmmode
    \ifx\testbx\f@series
      {\mathbf{#1\,\mathsc{#2}}}
    \else
      {\mathrm{#1\,\mathsc{#2}}}
    \fi
  \else
    \textup{#1\,{\mdseries\textsc{#2}}}%
  \fi
 }

\begin{document}

\lefthead{Modelling of the \ion{He}{ii} Transverse Proximity Effect}\righthead{T. M. Schmidt et al.}
\title{Modeling the \ion{He}{ii} Transverse Proximity Effect: Constraints on Quasar Lifetime and Obscuration}

\author{Tobias M. Schmidt\altaffilmark{1,2,6,*}, Joseph F. Hennawi\altaffilmark{1,2}, G\'abor Worseck\altaffilmark{2,3}, 
Frederick B. Davies\altaffilmark{1}, Zarija Luki{\'c}\altaffilmark{4}, Jose O{\~n}orbe\altaffilmark{5}}	
 
\altaffiltext{*}{e-mail: tschmidt@mpia.de}
\altaffiltext{1}{Department of Physics, University of California, Santa Barbara, CA 93106, USA}
\altaffiltext{2}{Max-Planck-Institut f\"ur Astronomie, K\"onigstuhl 17, D-69117 Heidelberg, Germany}
\altaffiltext{2}{Institut f\"ur Physik und Astronomie, Universit\"at Potsdam, Karl-Liebknecht Stra{\ss}e 24/25, D-14476 Golm, Germany}
\altaffiltext{4}{Lawrence Berkeley National Laboratory, CA 94720-8139, USA}
\altaffiltext{5}{Institute for Astronomy, Royal Observatory of Edinburg, Blackford Hill, Edinburgh EH9 3HJ, United Kingdom}
\altaffiltext{6}{Fellow of the International Max Planck Research School for Astronomy and
Cosmic Physics at the University of Heidelberg (IMPRS-HD).}

\begin{abstract}

The \ion{He}{ii} transverse proximity effect -- enhanced \ion{He}{ii} $\lya$ transmission in a background sightline caused by the ionizing radiation of a foreground quasar -- offers a unique opportunity to probe the emission properties of quasars, in particular the emission geometry (obscuration, beaming) and the quasar lifetime.
Building on the foreground quasar survey published in \citet{Schmidt2017}, we present a detailed model of the \ion{He}{ii} transverse proximity effect, specifically designed to include light travel time effects, finite quasar ages, and quasar obscuration.
We post-process outputs from a cosmological hydrodynamical simulation with a fluctuating \ion{He}{ii} UV background model, plus the added
effect of the  radiation from a single bright foreground quasar. 
We vary the age $\tTO$ and obscured sky fractions $\Oobsc$ of the foreground quasar, and explore the resulting effect on the \ion{He}{ii} transverse proximity effect signal. Fluctuations in IGM density and the UV background, as well as the unknown orientation of the foreground quasar, result in a large variance of the \ion{He}{ii} $\lya$ transmission along the background sightline. 
We develop a fully Bayesian statistical formalism to compare far UV \ion{He}{ii} $\lya$ transmission spectra of the background quasars to our models, and extract joint constraints on $\tTO$ and $\Oobsc$ for the six \citet{Schmidt2017} foreground quasars with the highest implied \ion{He}{ii} photoionization rates.
Our analysis suggests a bimodal distribution of quasar emission properties, whereby one foreground quasar, associated with a strong \ion{He}{ii} transmission spike, is relatively old ($22\Myr$) and unobscured $\Oobsc<35\%$, whereas three others are either younger than $10\Myr$ or highly obscured ($\Oobsc>70\%$).

\end{abstract}


\section{Introduction}

Active Galactic Nuclei (AGN) are the brightest non-transient sources in the universe and emit over a wide spectral range, in particular large amounts of ultraviolet (UV) ionizing radiation. However, after over 50 years of ongoing research (see \citealt{Padovani2017} for a recent review) many aspects of AGN are still poorly understood.  This involves especially the internal physical structure of AGN 
and the accretion and triggering process.
In this study we therefore try to constrain two key parameters of quasars, namely the geometry of their optical/UV radiation, and the duration of their emission episodes, which we will refer to as the quasar lifetime $t_{\rm Q}$. 
Besides their importance to the structure of AGN, these properties are relevant to studies of
galaxy evolution and AGN feedback, the growth history of supermassive black holes \citep{Soltan1982, Shankar2009, Kelly2010}, and in particular can have significant
implications for the thermal and ionization state of the intergalactic medium (IGM) \citep[e.g.][]{McQuinn2009, Khrykin2016, Khrykin2017, Chardin2017, Davies2017, LaPlante2017, DAloisio2017}.

The quasar lifetime is particularly uncertain and challenging to measure. Current constraints of $\tQ < 10^9\yr$ come from demographic arguments and the evolution of the AGN population, e.g. \citet{Martini2004} and references therein.  
Matches to the quasar luminosity function and clustering statistics deliver constraints on the duty cycle
$\approx 10^6 - 10^9\yr$ \citep{Adelberger2005, Croom2005, Shen2009, White2012, Conroy2013, LaPlante2015} but so far with large uncertainty due to the unknown way quasars populate dark matter halos.

The ionization state of diffuse gas in quasar environs provides a powerful technique to constrain the quasar lifetime. The best example is the well-known \ion{H}{i} $\lya$ line-of-sight proximity effect \citep{Carswell1982, Bajtlik1988, Scott2000, DallAglio2008b, Calverley2011}. Because it takes the IGM a finite time to respond to quasar radiation, the presence of a proximity effects sets a very robust but weak lower limit on the lifetime of $\tQ \gtrsim 10^4\yr$.
Stronger constraints can be derived at the highest redshifts $z\sim 6$ \citep[e.g.][]{Eilers2017} or from the analogous line-of-sight proximity effect in the \ion{He}{ii} Ly$\alpha$ forest \citep{Khrykin2016, Khrykin2017}. Lifetime constraints for individual quasars can be obtained by observing tracers of their ionizing radiation at substantial distances from the quasars. Examples are fluorescence of galaxies illuminated by quasars as claimed by \citet{Adelberger2006, Cantalupo2012, Trainor2013, Borisova2015}, the presence/absence of [\ion{O}{iii}] narrow emission line regions in AGN host galaxies \citep[$\tQ\sim10^5\,\mathrm{yr}$]{Schawinski2015}, the ionization state of metal absorption systems in quasar environs \citep[$\tQ>25\Myr$]{Goncalves2008},  or the \ion{He}{ii} transverse proximity effect \citep[$\tQ>10-25\Myr$]{Jakobsen2003, Worseck2006,Schmidt2017}, which we elaborate on in detail below.  So far, observations have not converged towards a common picture and theoretical investigations lack predictive power \citep[e.g][]{Springel2005, Hopkins2007, Novak2011}.

Another uncertainty is the emission geometry of quasars. There is clear evidence that quasars do not emit isotropically.
In the optical/UV regime, one observes a clear dichotomy in the spectral appearance of AGN which is in the \textit{unified model} \citep{Antonucci1993, Urri1995, Netzer2015} explained as a pure orientation effect. While in Type\,I quasars one has a direct view on the nuclear accretion disk and the broad line region, a dusty torus on parsec scales in the equatorial plane of the AGN completely or partially blocks the view on these regions in Type\,II AGN, leaving only the narrow line region observable. 
Within this framework, the fraction of quasars observed to be obscured (i.e. Type\,II quasars) $\fobsc$, is directly related to the fraction of the sky (solid angle) towards which each individual quasar is obscured $\Oobsc$. 
Current studies report obscured fractions in the range $\fobsc\sim 30\%-70\%$ \citep{Simpson2005, Brusa2010, Assef2013, Lusso2013, Buchner2015, Marchesi2016}, but with with substantial uncertainties.
It has been argued that the obscured fraction decreases strongly with luminosity
\citep[e.g.][]{Simpson1998, Honig2007, Assef2013}, which can be understood in the so-called \textit{receeding torus} model \citep{Lawrence1991}, although this has also been debated \citep{Lusso2013}. 
There are also other models favoring a different geometry for the dust distribution or a different mechanism for the obscuration \citep[e.g.][]{Elvis2000, Elitzur2006, Keating2012, Honig2017}. In addition, e.g. \citet{Buchner2017} attribute some of the obscuration to the host galaxy instead of a parsec scale torus. 
Other studies report observations incompatible with the idea of unification, e.g. \citet{Villarroel2017} or \citet{DiPompeo2017} claim intrinsic differences in the host galaxy properties of Type\,Is and Type\,IIs, which might point towards a different mechanism for the Type\,I / Type\,II dichotomy, possibly more in line with AGN evolution models suggested e.g. by \citet{Hopkins2007}.
These examples show, that the geometry of the UV emission of quasars is still highly uncertain. In particular, studies that directly constrain the emission geometry of individual AGN exist only for a few local Seyfert galaxies \citep[e.g.][]{Tadhunter1989, Wilson1993}.

The \ion{He}{ii} transverse proximity effect -- enhanced \ion{He}{ii} $\lya$ transmission along a background sightline caused by the ionizing radiation of a nearby foreground quasar -- offers a \textit{second view} on the foreground quasar and provides a unique opportunity to directly constrain the emission geometries of individual quasars.
While direct observations of the foreground quasar reveal its properties from Earth's vantage point, 
the observed \ion{He}{ii} $\lya$ transmission along the background sightline crucially depends on the emission of \ion{He}{ii} ionizing photons in (roughly) transverse direction.
In addition, these photons require time to reach the background sightline. Therefore, the \ion{He}{ii} $\lya$ transmission is, depending on the position along the background sightline, sensitive to the emission of the foreground quasar approximately one transverse light crossing time ago.
Hence, the transverse proximity effect is ideal to infer emission geometries and geometric constraints on quasar ages/lifetimes.
  
Different aspects and variations of this method based on \ion{H}{i} or \ion{He}{ii} $\lya$ absorption spectra have been described in detail e.g. by \citet{Dobrzycki1991, Smette2002, Adelberger2004, Visbal2008, Furlanetto2011}.
Notwithstanding the much smaller number of available sightlines, and the requirement for space based far-UV spectroscopic observations,
\ion{He}{ii} offers a substantial advantage compared to \ion{H}{i}. 
At intermediate redshifts ($z\sim3$) hydrogen is already highly ionized and the \ion{H}{i} $\lya$ opacity is low \citep[e.g.][]{GunnPeterson1965, Bolton2007c}. A foreground quasar therefore results in only a small  \ion{H}{i} transmission excess
in its surrounding proximity zone which may at least partly explain the frequent non-detections \citep[e.g.][]{Croft2004}.
In contrast, \ion{He}{ii} is believed to be reionized by quasars \citep{HaardtMadau2012, Khaire2017} which become frequent at relatively late cosmic times and represent a population of extremely bright but also rare sources. 
As such, before \ion{He}{ii} reionization is completed at $z\approx2.7$, the average \ion{He}{ii} photoionization rate is low, resulting in 
high \ion{He}{ii}~$\lya$ optical depths between $\tau_\mathrm{HeII}\approx1$ and $\tau_\mathrm{HeII}\gtrsim5$ \citep[e.g][]{Worseck2016, Davies2017}. 
In this regime, a single foreground quasar can produce a strong transmission enhancement in an otherwise nearly opaque background sightline.

Studies dedicated to the \ion{H}{i} transverse proximity effect \citep[e.g.][]{Liske2001, Schirber2004, Croft2004, Hennawi2006, Hennawi2007, Kirkman2008, Gallerani2008, Lau2016} did not lead to unambiguous  detections of the effect, in parts probably related to its relative weakness at intermediate redshifts and the cosmic overdensities hosting quasars which counteract the enhanced photoionization.
For \ion{He}{ii} however, \citet{Jakobsen2003} established the picture of the transverse proximity effect by the detection of a foreground quasar associated to a strong \ion{He}{ii} transmission spike in the spectrum of Q\,0302-003 \citep{Heap2000}. However, despite the striking nature of this discovery, further progress in this field was limited by the lack of additional associations of this type. 

In \citet{Schmidt2017} we presented the results of a dedicated foreground quasar survey targeting the vicinity of 22 \ion{He}{ii} sightlines. We substantially expanded the sample of known foreground quasars to a set of 20 quasars that should cause a \ion{He}{ii} photoionization rate $>2\times10^{-15}\,\mathrm{s}^{-1}$ at the background sightline, sufficient to cause a transverse proximity effect. This allowed for the first time a statistical analysis of the effect.
Stacking the background spectra at the foreground quasar locations,  we found statistical evidence for the \ion{He}{ii} transverse proximity effect and derived a heuristic constraint on quasar lifetime of $\tQ > 25\Myr$. 
However, among the four foreground quasars with the highest estimated \ion{He}{ii} photoionization rates at the background sightline, only the previously known prototype object along the Q\,0302-003 sightline \citep{Heap2000,Jakobsen2003} showed a strong \ion{He}{ii} transmission spike. 
Surprisingly, all three newly discovered foreground quasars -- despite having higher estimated \ion{He}{ii} photoionization rates and exceeding the \ion{He}{ii} UV background by an order of magnitude -- exhibit very low transmission or even saturated absorption along the background sightline.
In \citet{Schmidt2017} we therefore speculated that either short quasar emission episodes or a high level of obscuration is required to explain these three objects.
However, there existed at that time no quantitative prediction for the appearance of the \ion{He}{ii} spectra in the vicinity of these quasars, in particular not encompassing IGM stochasticity, finite quasar ages and obscuration.

In this work, we therefore follow up on our previous study with a detailed modeling of the expected \ion{He}{ii} transmission signal, focussing on the six foreground quasars with the highest \ion{He}{ii} photoionization rate from \citet{Schmidt2017}. We use outputs from the \Nyx cosmological hydrodynamical simulations \citep{Almgren2013, Lukic2015} and post-process these with a photoionization model composed of the radiation from a single, bright foreground quasar on top of a semi-numerical, fluctuating \ion{He}{ii} UV background model \citep{Davies2017}.
For the foreground quasar, we vary the quasar age $\tTO$ and obscuration $\Oobsc$ and explore the combined effect for the \ion{He}{ii} transverse proximity effect.
To embrace the stochastic nature of quasar orientation, \ion{He}{ii} UV background fluctuations  and IGM density structure, we compute many Monte Carlo realizations, allowing us to quantify for the first time the expected amount of fluctuations in observations of the \ion{He}{ii} transverse proximity effect.
Using a fully Bayesian statistical approach, we compare our specifically designed models to the observed \ion{He}{ii} spectra and infer joint probabilities for quasar ages $\tTO$ and obscured sky fractions $\Oobsc$ of the six individual quasars.

This paper is structured as follows. In \S~\ref{Sec:Sample} we summarize the subset of foreground quasars and \ion{He}{ii} sightlines from \citet{Schmidt2017} that we model. The computation of our models, starting from outputs of the cosmological hydrodynamical simulation, application of UV background, and quasar emission models and the calculation of the final mock spectra are described in \S~\ref{Sec:3}. The statistical approach developed for the comparison of the models to the \ion{He}{ii} observations is described in \S~\ref{Sec:4}. We derive joint the probability distribution of $\tTO$ and $\Oobsc$ in \S~\ref{Sec:Results}, and discuss the implications of our measurements in \S~\ref{Sec:Discussion}.

Throughout the paper we use a flat $\Lambda$CDM cosmology with $H\mathrm{_0 =68.5\,km \: s^{-1} \: Mpc^{-1}}$, $\mathrm{\Omega_\Lambda = 0.7}$, $\mathrm{\Omega_m = 0.3}$ and $\mathrm{\Omega_b = 0.047	}$ which was used for the computation of the \Nyx hydro simulation and is broadly consistent with the \citet{Planck2015} results. We use comoving distances and denote the corresponding units as $\cMpc$.
For most of the paper (except \S~\ref{Sec:Non-Lightbulb}) we consider a simple lightbulb model for the quasar lightcurve in which the quasar turns on, shines with constant luminosity for its full lifetime $\tQ$ until it turns off. This timespan is however different from the quasar age $\tTO$, which describes the time from turning on until emission of the photons that are observed on Earth today.
Magnitudes are given in the AB system.

\section{Data Sample}
\label{Sec:Sample}

We use the sample of \ion{He}{ii} sightlines and foreground quasars from \citet{Schmidt2017}. However, we restrict our analysis to the six foreground quasars with the highest \ion{He}{ii} photoionization rate at the background sightline (see \S~\ref{Sec:Quasar_Ionization_Model} for a formal definition), and therefore strongest expected transverse proximity effect signal. Under the assumption of isotropic emission and infinite quasar lifetime, these quasars should cause a peak ionization rate at the background sightline between $7.3$ and $19\times10^{-15}\,\mathrm{s}^{-1}$, and therefore exceed the expected intergalactic \ion{He}{ii} UV background \citep{Faucher-Giguere2009, HaardtMadau2012, Khrykin2016, Khaire2017} by approximately one order of magnitude. An overview of the objects studied is given in Table~\ref{Tab:Objects} and Far UV spectra of all six \ion{He}{ii} sightlines are shown in \S~\ref{Sec:Results}, Figure~\ref{Fig:Key_Plot}.

\begin{deluxetable*}{lrrrrrrrr}
\tablecolumns{9}
\tablewidth{0.95\linewidth}
\tablecaption{
Key properties of the foreground quasars used for this study. 
}
\tablehead{
\colhead{\ion{He}{ii} Sightline}	& \colhead{$RA$ (2000)}	& \colhead{$Dec$ (2000)} 	& \colhead{$z$}	& \colhead{$r$}	& \colhead{$M_{1450}$}	& \colhead{$\Delta\theta$}	& \colhead{$D_\mathrm{prop}$}	& \colhead{$\Gamma_\mathrm{QSO,\:max}^\mathrm{HeII}$}	\\
\colhead{}				& \colhead{degree}	& \colhead{degree}      	& \colhead{}	& \colhead{mag}	&	\colhead{mag}   	& \colhead{arcmin}      	& \colhead{$\pMpc$}     	& \colhead{$\mathrm{s}^{-1}$}		
}
\startdata
HE2QS\,J2149$-$0859     & $327.23032$    & $  -9.02613$    & $2.815$    & $19.3$    & $-26.0$    & $ 8.4$    & $ 4.0$    & $1.91\e{-14}$\\
HE2QS\,J0916$+$2408     & $139.16456$    & $ +24.19545$    & $2.846$    & $21.0$    & $-24.3$    & $ 5.6$    & $ 2.7$    & $9.44\e{-15}$\\
SDSS\,J1101$+$1053      & $165.51796$    & $ +10.95631$    & $2.912$    & $20.8$    & $-24.5$    & $ 4.8$    & $ 2.3$    & $1.60\e{-14}$\\
HS\,1157$+$3143         & $180.41579$    & $ +31.59376$    & $2.917$    & $18.4$    & $-27.0$    & $21.9$    & $10.4$    & $7.36\e{-15}$\\
Q\,0302$-$003           & $ 46.14721$    & $  -0.04750$    & $3.050$    & $20.6$    & $-24.9$    & $ 6.5$    & $ 3.0$    & $1.23\e{-14}$\\
SDSS\,J1253$+$6817      & $193.87605$    & $ +68.33807$    & $3.210$    & $19.4$    & $-26.1$    & $ 9.4$    & $ 4.4$    & $1.90\e{-14}$
\enddata

\label{Tab:Objects}
\end{deluxetable*}
 
Despite their high peak photoionization rates at the background sightline, we observe no strong \ion{He}{ii} transverse proximity effect for most of these foreground quasars. Only the prototype object at redshift $z=3.05$ close to the Q\,0302$-$003 sightline is associated with a strong \ion{He}{ii} transmission peak \citep{Heap2000, Jakobsen2003}. The others show ordinary and sometimes saturated \ion{He}{ii} absorption. We have discussed the absence of transmission spikes for three of the four strongest foreground quasars in \citet{Schmidt2017}.

For this study, we include two other objects with high \ion{He}{ii} ionization rates, one at  $z=2.846$ along the HE2QS\,J0916$+$2408 sightline and one close to HS\,1157$+$3143 at $z=2.917$. Owing to its lower redshift, the HE2QS\,J0916$+$2408 sightline shows in general higher \ion{He}{ii} transmission around 20\%, with a broad transmission structure that might be associated with the foreground quasar, or could just be a random UV background fluctuation. HS\,1157$+$3143 shows low \ion{He}{ii} transmission around 8\% with a very subtle broad bump around the foreground quasar position.
There exists another foreground quasar with comparably high ionization rate along the SDSS\,J1253$+$6817 sightline at $z=2.904$. However, we have incomplete far UV coverage along the background sightline and therefore do not include this object.

\label{Sec:EnhancementStatistic}
In \citet{Schmidt2017} we quantified the strength of the observed transverse proximity effect by introducing the transmission enhancement statistic. We therefore measure the average \ion{He}{ii} transmission in a $\pm15\cMpc$ wide window around the foreground quasar and compare this with the average transmission outside this window. In this study we continue using this statistic which is formally defined as
\begin{equation}
\xiTPE = \langle \F_{|R_\parallel|<15\cMpc}\rangle - \langle \F_{15\cMpc<|R_\parallel|<65\cMpc}\rangle
\label{Eq:xiTPE},
\end{equation}
where $F$ denotes the \ion{He}{ii} transmission and $R_\parallel$ the coordinate along the background sightline.
In contrast to \citet{Schmidt2017} we reduce the extent over which the background transmission is measured from $\pm120\cMpc$ to $\pm65\cMpc$ since the \Nyx simulation box used for this study only offers a pathlength of $146\cMpc$. 

Using $\xiTPE$ instead of simply the average \ion{He}{ii} transmission $\langle \F_{|R_\parallel|<15\cMpc}\rangle$ has the advantage that, to first order, the dependence on the \ion{He}{ii} UV background is removed.  It thus better isolates the effect of the foreground quasar from unassociated  background fluctuations. This is for instance illustrated in Figure~\ref{Fig:Test_UVB} in \S~\ref{Sec:UV_Background}
.

\section{Models / Simulations}
\label{Sec:3}

For this work we take outputs from a cosmological hydrodynamical simulation and post-process them with a photoionization model.
This photoionization model is composed of the radiation from a single, bright foreground quasar and a fluctuating \ion{He}{ii} UV background from \citet{Davies2017}.
In the following, we first line out the extraction of skewers from the simulation box, the general calculation of \ion{H}{i} and \ion{He}{ii} ionization states and the computation of mock spectra. 
We then present in more detail the fluctuating \ion{He}{ii} UV background model and the calibration procedure to make this model match the \ion{He}	{ii} observations. Finally, we describe our model for the quasar ionization radiation, including the effects of finite quasar age and quasar obscuration.

\subsection{\Nyx Cosmological Hydrodynamical Simulations}

We use simulations computed with the  Eulerian hydrodynamical simulation code \Nyx \citep{Almgren2013, Lukic2015}. 
The simulation box has a large size of $100\,h^{-1}\cMpc$ which is required to capture the full extent of a bright quasars \ion{He}{ii} proximity zone. The hydrodynamics is computed on a fixed grid of $4096^3$ resolution elements and the same number of dark matter particles are used for computation of the gravitational field. This results in a resolution of $36\ckpc$ per pixel, required and sufficient to resolve the \ion{H}{i} \citep{Lukic2015} and \ion{He}{ii} $\lya$~forest. 
The simulation runs make no use of adaptive mesh refinement since the \ion{H}{i}~$\lya$~forest signal originates from the majority of the volume \citep{Lukic2015} and the \ion{He}{ii} signal actually stems from the underdense regions \citep[e.g][]{Croft1997}. Refining the resolution in the dense regions at the expense of underdense regions is therefore not beneficial for our case.
Also, since the prime objective of the simulation is IGM science, no star or galaxy formation prescriptions was included. 
The simulation was run using a homogeneous, optically thin UV background with photoionization and heating rates from \citet{HaardtMadau2012}. As described below, we rescale the \ion{H}{i} and \ion{He}{ii} photoionization rates to closely match observations but keep the thermal structure unchanged. 

We use the density, velocity and temperature fields of a single simulation output at $z=3$ and extract skewers that will be post-processed to simulate the observed \ion{He}{ii} Ly$\alpha$ transmission along the background sightlines. 
We tailor these to match our data sample as closely as possible, in particular we create for each foreground quasar in our sample a set of skewers with matched transverse separation, redshift and quasar luminosity.  
We center the foreground quasars on $\approx 10^{12}\Msun$ halos, the preferred mass of AGN halos \citep[e.g][]{White2012}.
As described in more detail in \citet{Sorini2017}, halos in the \Nyx simulations are identified by finding topologically connected components above  138$\times$ mean density (Luki{\'c} et al. in prep.). This gives similar results than the particle-based friends-of-friends algorithm \citep{Davis1985}. 
From the \Nyx halo catalog we select for each model e.g. the $2500$ halos with mass closest to $10^{12}\Msun$ and from this set randomly reject 90\% to avoid deterministic behavior. 

Given the list of selected halos, skewers are extracted along one of the grid axes with a transverse offset from the halo center matched to the observed separation between the foreground quasar and background sightline. The position angle between halo and skewer is randomly chosen. Multiple, skewers (e.g. 20) are extracted around each halo.
Along the line of sight, we center the skewer on the halo position in redshift space, taking the peculiar velocity of the halo into account. With the observed redshift of the foreground quasar as the origin, we assign individual redshifts to every pixel of the skewer. 
To better represent redshift evolution of the density field along the sightline we rescale the density of each pixel
accordingly
\begin{equation}
\rho(z) = \rho_\mathrm{sim}  \times \left( \frac{z+1}{z_\mathrm{sim}+1} \right) ^3\:.
\end{equation}
However, since the relevant range in redshift only spans $2.75 < z < 3.25$ (see Table~\ref{Tab:Objects}), this correction is small.
We convert from simulated cosmic baryon density to hydrogen and helium number density $n_\mathrm{H}$ and $n_\mathrm{He}$ using the primordial abundances of these elements, 76\% and 24\% \citep{Coc2015}.  
The temperature and velocity field are taken directly from the simulation box without any change.

\subsection{Ionization State for Hydrogen and Helium}

After extracting temperature $T$, velocity and cosmic baryon density from the Nyx simulation box and converting these to $n_\mathrm{H}$ and $n_\mathrm{He}$ assuming primordial abundances, we solve for the ionization state of hydrogen and helium. 
This requires a description of the corresponding photoionization rates $\Gamma_\mathrm{tot}^\mathrm{HI}$ and $\Gamma_\mathrm{tot}^\mathrm{HeII}$. Related to the presence of a bright foreground quasar and due to the fluctuating \ion{He}{ii} UV background model,  \ion{H}{i} and \ion{He}{ii} photoionization rates are spatially variable along our skewers. A detailed description of the adopted photoionization model follows later in \S~\ref{Sec:UV_Background} and \S~\ref{Sec:Quasar_Ionization_Model}.

We assume ionization equilibrium and ignore time-evolution and non-equilibrium effects. The equilibration timescale for \ion{He}{ii} is rather large, depending on the \ion{He}{ii} photoionization rate of the order of a few million years. However, the timescales our sightline geometries are sensitive to are even longer.  We discuss non-equilibrium effects and more complicated quasar lightcurves in \S~\ref{Sec:Non-Lightbulb}.

Within the regime we are operating, around $z\sim3$, hydrogen reionization as well as \ion{He}{i} reionization is completed and all hydrogen in the IGM is highly ionized \citep[e.g.][]{HaardtMadau2012, Planck2016}. 
We thus can separate the calculation of hydrogen and helium ionization state and avoid solving a coupled problem.
In a first step, we calculate the hydrogen ionization state, ignoring the \ion{He}{ii}~$\rightarrow$~\ion{He}{iii} transition, i.e. assuming $n_\mathrm{HeIII} = 0$. We follow the general approach as it is described e.g. in \citet{Rahmati2013}. Ionization equilibrium is expressed by
\begin{equation}
n_\mathrm{HI} \, \Gamma_\mathrm{tot}^\mathrm{HI} = \alpha_\mathrm{A}^\mathrm{HII} \, n_{e^-} \, n_\mathrm{HII}
\label{Eq:HI}
\end{equation}
with $n_\mathrm{HI}$, $n_\mathrm{HII}$ and $n_{e^-}$ are the number densities of neutral hydrogen, ionized hydrogen and free electrons, respectively. 
The ionization rate $\Gamma_\mathrm{tot}^\mathrm{HI}$ is the sum of photoionization $\Gamma_\mathrm{phot}^\mathrm{HI} = \GuvbH + \GqsoH$ and collisional ionization. For the photoionization we include the self-shielding prescription from \citet{Rahmati2013} in which the effective photoionization rate in high-density regions with $n_\mathrm{H} \gtrsim 5 \times 10^{-3}\,\mathrm{cm}^3$ is substantially reduced.
For collisional ionization we assume $\Gamma_\mathrm{col}^\mathrm{HI} = \Lambda^\mathrm{HI} \, n_{e^-}$ with 

\begin{equation}
\Lambda^\mathrm{HI}(T) = 1.17^{-10} \; \frac{  {(T/\mathrm{K})}^{1/2} \; e^{ -157809\,\mathrm{K} \: / \: T } }{ 1 + \sqrt{ \, T \: / \: 10^5\,\mathrm{K} \,} }  \, \mathrm{cm^3} \, \mathrm{s^{-1}}
\end{equation}
from \citet{Theuns1998}. 
We tie the fraction of helium in the \ion{He}{i} and \ion{He}{ii} states to the hydrogen ionization state by simply assuming $n_\mathrm{HeII}/n_\mathrm{He} = n_\mathrm{HII}/n_\mathrm{H}$.
Given the similar ionization energies, this is justified and a common assumption. 
The electron density in Equation~\ref{Eq:HI} therefore has to be $n_{e^-} = n_\mathrm{HII} \; c_{e^-}^\mathrm{HII}$
with $c_{e^-}^\mathrm{HII} = 1.079$ being a correction factor that accounts for the electrons contributed by the singly ionization of helium at the level of the cosmic primordial mass fractions of hydrogen and helium.
For $\alpha_\mathrm{A}^\mathrm{HI}(T)$ we use the Case~A recombination coefficients from \citet{Storey1995}.
With these inputs, Equation~\ref{Eq:HI} becomes a simple quadratic equation that can be easily solved for the hydrogen ionized fraction.

In the second step, we compute the number densities of singly (\ion{He}{ii}) and doubly ionized helium (\ion{He}{iii}),
which depend on the hydrogen ionization state. For this calculation  we ignore \ion{He}{i} and assume that all helium is at least singly ionized ($n_\mathrm{HeII} + n_\mathrm{HeIII} = n_\mathrm{He}$) which is an excellent approximation given that $n_{\rm HeI}\slash n_{\rm HeII}\sim 10^{-5}$. In complete analogy to Equation~\ref{Eq:HI} helium ionization equilibrium is expressed as
\begin{equation}
n_\mathrm{HeII} \, \Gamma_\mathrm{phot}^\mathrm{HeII} = \alpha_\mathrm{A}^\mathrm{HeIII} \, n_{e^-} \, n_\mathrm{HeIII}\:.
\label{Eq:HeII}
\end{equation}
For \ion{He}{ii} we do not include collisional ionization or self-shielding corrections. We again use Case~A recombination coefficients $\alpha_\mathrm{A}^\mathrm{HeII}(T)$ from \citet{Storey1995}.
The electron density is now dominated by the electrons supplied by ionized hydrogen:
\begin{equation}
n_\mathrm{e^-} = n_\mathrm{HII} \, c_{e^-}^\mathrm{HII} + n_\mathrm{HeIII} \: .
\end{equation}
This is the reason $n_\mathrm{HII}$ had to be computed  a priori. With all required information collected, Equation~\ref{Eq:HeII} can be solved for $n_\mathrm{HeIII}$.

The additional electrons released by the \ion{He}{ii}~$\rightarrow$~\ion{He}{iii} transition in principle effect the hydrogen ionization state. The correct way would be to iterate over Equation~\ref{Eq:HI} and \ref{Eq:HeII} until convergence. However, the total effect on $n_{e^-}$ is small ($<8\%$) and has for highly ionized hydrogen a totally negligible impact on $n_\mathrm{HII}$ ($<10^{-6}$) and therefore on $n_\mathrm{HeII}$, completely insignificant compared to the uncertainties in the UV background and effective optical depth measurement. This justifies solving hydrogen and helium ionization state independent of each other.

\subsection{Computing Synthetic Spectra}

After determining $n_\mathrm{HI}$ and $n_\mathrm{HeII}$ along the skewers as stated above, the final step in our modeling procedure is to create synthetic spectra.
For each pixel along the skewers we compute an individual Voigt absorption line profile with appropriate strength, line width and velocity shift corresponding to the physical conditions in that pixel. Oscillator strengths are taken from \citet{Verner1996a}.
We benefit here from the high resolution of the Nyx box ($36\ckpc$ or $\mathrm{2.8\,km\,s^{-1}}$) which is sufficient to resolve \ion{H}{i} and \ion{He}{ii} $\lya$~forests ($\mathrm{\approx7.6\,km\,s^{-1}}$ and $\mathrm{3.8\,km\,s^{-1}}$).
Redshift space distortions (peculiar velocities) are included by displacing the absorption profile with the line of sight velocity from the \Nyx simulation.
Thermal broadening is computed according to $\sigma_\mathrm{th} = \sqrt{ \frac{k_\mathrm{B} \, T}{m_\mathrm{ION}} }$ for the Doppler broadening%
\footnote{This describes the standard deviation of the Gaussian part of the Voigt profile. The often used \textit{Doppler Parameter} is $b_\mathrm{th} = \sqrt{2} \times \sigma_\mathrm{th}$.} with $T$ denoting the gas temperature in a pixel and $m_\mathrm{ION}$ the atomic masses of hydrogen or helium. The Lorentzian scale parameter is based on the transition probability from \citet{Verner1996a}. 
The final transmission spectrum at a pixel in redshift space is the combination of all the absorption profiles along the skewer.  We do not convolve the spectra with any instrumental line-spread function since the measurements are obtained in at least $16\cMpc$ wide bins, much broader than the typical $\approx2\cMpc$ resolution of the \ion{He}{ii} spectra. 

Using the velocity structure from the hydrodynamical simulation is extremely important. In most cases, significant \ion{He}{ii} transmission stems predominantly from underdense regions. In these voids, the velocity field is usually divergent, making them appear larger in redshift space which leads to a \ion{He}{ii} mean transmission e.g. $3\times$ higher for $\GuvbHe = 10^{-15}\,\mathrm{s}^{-1}$ compared to the case without peculiar velocities.

\subsection{\ion{H}{i} and \ion{He}{ii} UV Background}
\label{Sec:UV_Background}

To obtain realistic \ion{He}{ii} transmission spectra, in particular in the absence of a foreground quasar, we have to rely on models for the corresponding UV backgrounds. \citet{Onorbe2017} obtained an empirical fit for the cosmic mean transmitted \ion{H}{i} flux $\langle F_\mathrm{HI} \rangle$ to existing measurements \citep{Fan2006, Becker2007, Kirkman2007, Faucher-Giguere2008, Becker2013} of the form
\begin{equation}
 \tau_\mathrm{HI} = 0.00126 \times  e^{ 3.294 \, \times \sqrt{z} }
\end{equation}
where $\tau_\mathrm{HI} = \ln{ \langle F_\mathrm{HI} \rangle }$ denotes the effective optical depth and $z$ the redshift.
For simulation snapshot available at $z= 2.0, 2.2, 3.0, 3.5, 4.0$ 
we measure the mean transmission in a large set of random skewers and iteratively adjust the homogeneous \ion{H}{i}~UV background until the mean transmission matches the fit from \citet{Onorbe2017}. We interpolate these $\GuvbH$ values determined for the fixed redshifts using a cubic spline to obtain a smooth function $\GuvbH(z)$. This allows us to assign the appropriate \ion{H}{i} UV background matched to the redshift of each pixel. 

\label{Sec:HeII_UVB}
Obtaining the correct \ion{He}{ii} UV background poses a bigger challenge. For redshifts $z>2.7$ helium reionization is incomplete and no homogeneous UV background has formed yet. Instead, the metagalictic \ion{He}{ii} ionization field is patchy and fluctuating \citep[e.g.][]{McQuinn2009, Worseck2016, Davies2017}. Without using the correct ionizing background that includes these fluctuations we can not expect to obtain realistic models for the effect of individual quasars. We therefore use the fluctuating UV background model from \citet{Davies2017} and add on top of that the ionizing radiation of the foreground quasars (\S~\ref{Sec:Quasar_Ionization_Model}).

The adopted approach is clearly a simplification. 
However, solving the full \ion{He}{ii} reionization history using self-consistent radiative transfer hydrodynamical calculations in a cosmological volume at high resolution including a statistical population of quasars matched to a given quasar luminosity functions and at the same time including the sample of explicitly observed quasars along the \ion{He}{ii} sightlines with variation and inference of quasar emission properties ($\tTO$, $\Oobsc$, etc) is infeasible with current methods.
We therefore have to investigate the effect of single isolated foreground quasars decoupled from the surrounding UV background. There might be by-chance proximity regions of observed or unobserved, Type\,I, Type\,II or even extinct quasars in the vicinity of the foreground quasar we focus on and it is impossible to model these explicitly. However, using the \citet{Davies2017} fluctuating \ion{He}{ii} UV background takes at least to some degree care of this since this UV background model is based on the combined and overlapping effect of proximity regions around a realistic quasar population. In addition, we only focus on the foreground quasars with the highest \ion{He}{ii} ionization rates at the background sightline which dominate over the \ion{He}{ii} UV background by approximately one order of magnitude. This makes our analysis less dependent on the exact details of the adopted \ion{He}{ii} UV background model.

\subsubsection{Fluctuating \ion{He}{ii} UV Background Model}

\begin{figure*}
 \begin{center}
  \includegraphics[width=0.85\textwidth]{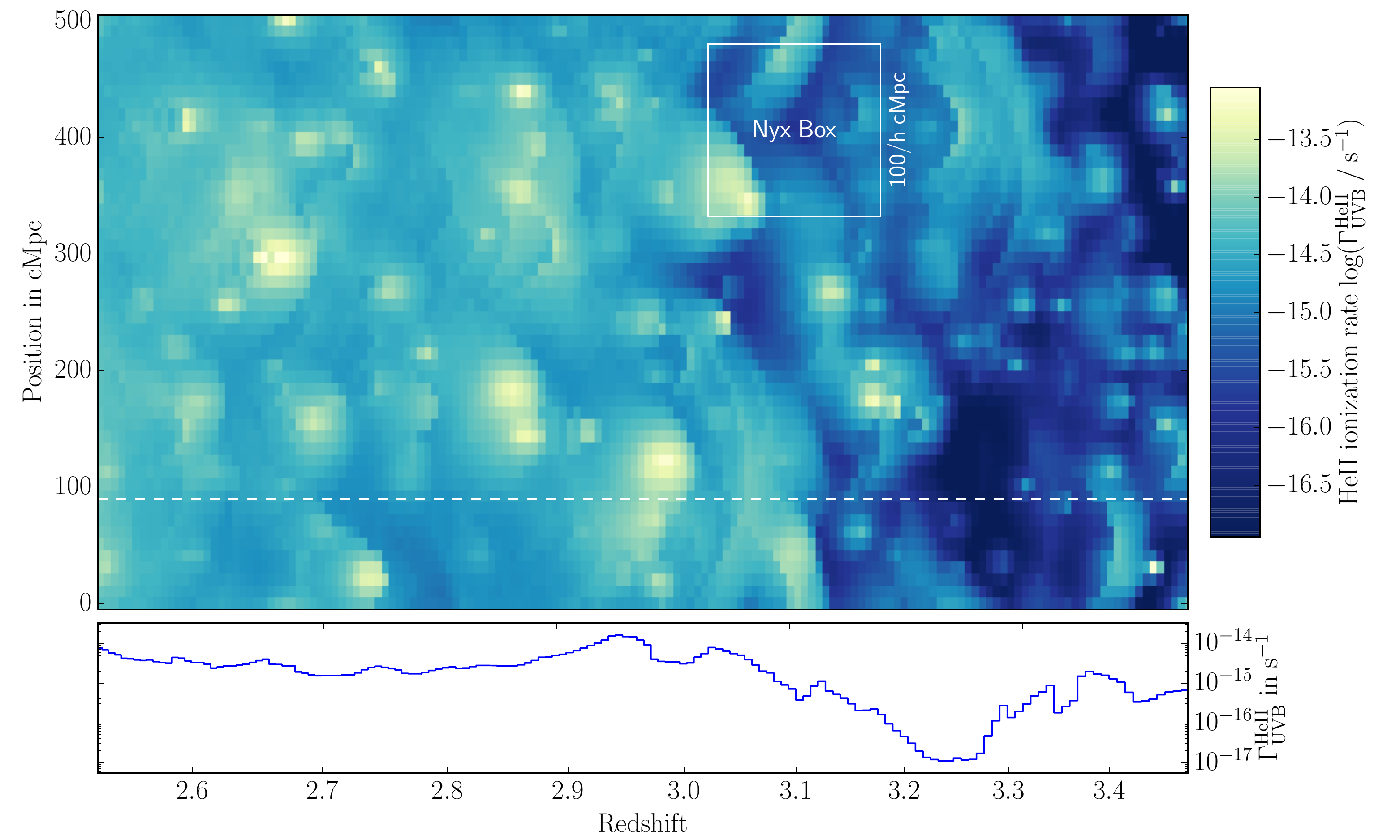}
  \caption{Visualization of the semi-analytic \ion{He}{ii} UV background model from \citet{Davies2017}. The top panel shows the \ion{He}{ii} ionization rate along a slice through the box in lightcone projection, meaning the vertical axis represents spatial position and the horizontal axis indicates position in redshift space as it appears for an observer on Earth. Clearly visible are the parabolic ionization regions around individual quasars. The size of our high-resolution \Nyx box is indicated. The bottom panel shows $\GuvbHe(z)$ along the dashed skewer.}
  \label{Fig:FredUVB}
 \end{center}
\end{figure*}

The \citet{Davies2017} \ion{He}{ii} UV background model is based on a large $500\cMpc$ box, sampled with $10\cMpc$ spatial resolution in which explicit sources of \ion{He}{ii} ionizing photons are statistically placed according to the \citet{Hopkins2007} quasar luminosity function. Each of these sources emits isotropically for a time span of $50\Myr$ and their radiation is propagated using a 3D radiative-transfer calculating with finite speed of light.
The calculation includes an explicit treatment of a spatially varying \ion{He}{ii} mean free path computed self-consistently under the assumption of local photoionization equilibrium. Figure \ref{Fig:FredUVB} shows a lightcone projection of the \ion{He}{ii} ionization rate $\GuvbHe$ along a random slice through the simulation volume. 
We calculate \ion{He}{ii} background photoionization rates along our skewers by randomly drawing $\GuvbHe$ lightcone skewers from the \citet{Davies2017} box (sampled on $\approx6\cMpc$ pixels in the redshift direction)  and interpolate these to the higher resolution of the \Nyx box using a cubic spline interpolation. This naturally includes the redshift evolution of the \ion{He}{ii} UV background along our sightlines. See lower panel of Figure~\ref{Fig:FredUVB} for an example.

\subsubsection{Calibration of the \ion{He}{ii} UV background model}

\begin{figure*}
 \begin{center}
  \includegraphics[width=0.95\textwidth]{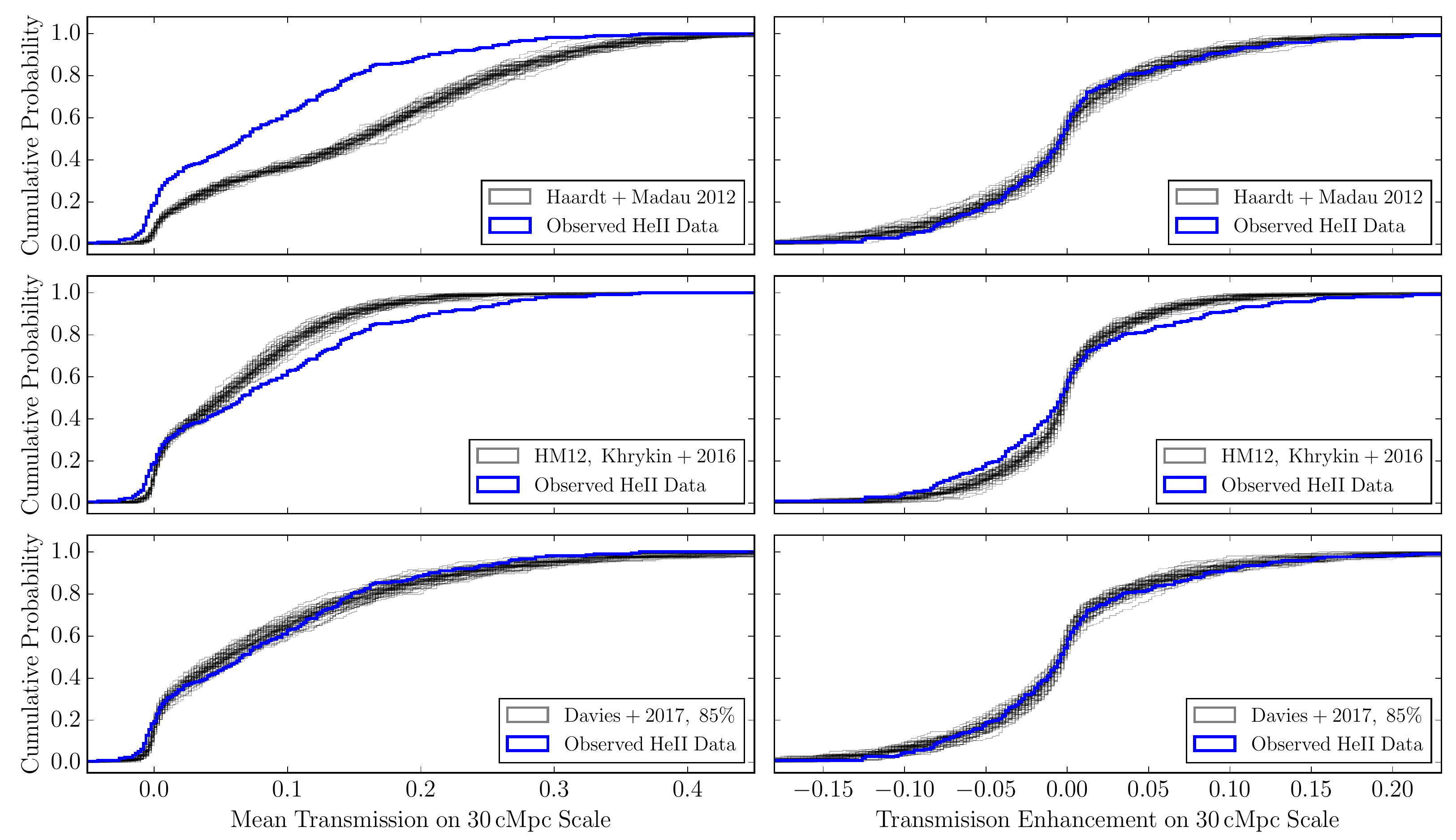}
  \caption{
   Cumulative histograms of the \ion{He}{ii} transmission (\textit{left}) and transmission enhancement (\textit{right}) measured in $30\cMpc$ wide bins along all 22 available \ion{He}{ii} sightlines. The blue curves show the actual data, black represent 50 random model realizations. The top row shows models using the homogeneous \citet{HaardtMadau2012} \ion{He}{ii} UV background model. This obviously produces too high transmission values. The central row therefore shows a case in which we rescale the UV background to $\GuvbHe=10^{-14.9}\,\mathrm{s}^{-1}$ at $z=3.1$ as found by \citet{Khrykin2016}. Here, the models produce a too narrow distribution for the transmission enhancement statistic. Excellent match in both statistics is achieved using the \citet{Davies2017} fluctuating  \ion{He}{ii} UV background model (\textit{bottom row}).
  }
 \label{Fig:Test_UVB}
 \end{center}
\end{figure*}

As pointed out above, adopting the correct \ion{He}{ii} UV background including the right amount of fluctuations is absolutely crucial in the context of this study. We therefore thoroughly test different UV background schemes and calibrate the adopted model to make sure it reproduces existing \ion{He}{ii} observations.

For this quantitative comparison, we use the full \ion{He}{ii} dataset from \citet{Worseck2016} and \citet{Schmidt2017}, composed of 22 \ion{He}{ii} sightlines. These sightlines represent a random sample and were selected independent of any possible foreground quasars. We  measure the \ion{He}{ii} transmission as well as the transmission enhancement $\xiTPE$ (for a definition see Equation~\ref{Eq:xiTPE}) in consecutive $30\cMpc$ wide bins along these sightlines. For each of the two statistics we obtain 212 measurements, considering only those bins that have full spectral coverage. 
To compare these measurements with \ion{He}{ii} UV background models, we compute for each of the 212 bins a large set of skewers centered on the same redshift and measure \ion{He}{ii} transmission and transmission enhancement in the simulated \ion{He}{ii} spectra. We measure the noise in the data and add this to our models as described in \S~\ref{Sec:NoiseMeasurement}. 

The result is presented in Figure~\ref{Fig:Test_UVB}. We show the cumulative histogram of the 212 measurements (blue) and 50 independent random realizations of the modeled \ion{He}{ii} dataset (black lines) for each of the three analyzed \ion{He}{ii} UV background models.
As shown in the top row, using the \citet{HaardtMadau2012} \ion{He}{ii} UV background leads to a substantially higher \ion{He}{ii} transmission than seen in the observations (left panel). However, the transmission enhancement $\xiTPE$, being a differential measurement in nature, is far less susceptible to the absolute level of the \ion{He}{ii} mean transmission and yields a far better match to the data than the transmission statistic itself (right panel).

The middle row shows the same approach, but rescaling the \citet{HaardtMadau2012} UV background to $\GuvbHe=10^{-14.9}\,\mathrm{s}^{-1}$ at $z=3.1$ as found by \citet{Khrykin2016}.
Now, the model (black lines) shows too few high-transmission regions relative to the data (left panel, blue histogram).
Also, it does not create enough fluctuation in the transmission enhancement (right panel), since the cumulative $\xiTPE$ probability distribution of the models is too steep to match the data (central right panel of Figure~\ref{Fig:Test_UVB}).

In contrast to these homogeneous UV background models, we achieve, as presented in the bottom row of Figure~\ref{Fig:Test_UVB}, excellent agreement between model and data for the flux statistic as well as for the transmission enhancement using the fluctuating \ion{He}{ii} UV background model from \citet{Davies2017}. We found the best match when rescaling the \citet{Davies2017} UV background to 85\% amplitude. This rescaling is well within the uncertainties of the model, which
was not tuned to match any particular \ion{He}{ii} transmission level. 

This test indeed compares the correct quantities. Our observed \ion{He}{ii} spectra do show signatures of quasar proximity zones but so do the $\GuvbHe$ skewers from the \citet{Davies2017} UV background model. In both cases we measure \ion{He}{ii} transmission at random positions which are uncorrelated to possible foreground quasars. 
The excellent agreement shows that our mildly rescaled \citet{Davies2017} fluctuating UV background model is actually capable of reproducing the observed \ion{He}{ii} transmission properties at random positions along the \ion{He}{ii} sightlines. This allows us to proceed by adding the ionizing radiation of individual quasars on top of the UV background to calculate transmission profiles that will finally be compared to \ion{He}{ii} spectra in the vicinity of the bright foreground quasars that we consider in detail.

\subsection{Modeling Foreground Quasar Emission and Ionization Rates}
\label{Sec:Quasar_Ionization_Model}

In a first step, we calculate the \ion{H}{i} and \ion{He}{ii} ionizing fluxes for positions along the background sightlines, given that they are illuminated by of the foreground quasars. This is identical to the approach outlined in \citet{Schmidt2017}.
The exact conditions under which point might not be illuminated due to obscuration or finite quasar age will be discussed in \S~\ref{Sec:QuasarObscuration} and \S~\ref{Sec:QuasarLifetime}    

Based on the $r$-band magnitude and the \citet{Lusso2015} quasar template we compute $M_{1450}$ and the quasar luminosity $L_\nu$.
Conversion to flux density $F_\nu$ at the background sightline is done according to
\begin{equation}
F_\nu = L_\nu \; \frac{1}{4\pi\:D_\mathrm{prop}^2}\;e^{-\frac{D_\mathrm{prop}}{\lambda_\mathrm{mfp}}}\;.
\label{Eq:GeometricDilution}
\end{equation}
Here, $D_\mathrm{prop}$ denotes the proper 3-D distance from the foreground quasar to a specific position at the background sightline and $\lambda_\mathrm{mfp}$ is the mean free path to \ion{He}{ii} ionizing photons.
Since the separations we deal with in our analysis are moderate ($D_\mathrm{prop}\lesssim 6.5\pMpc$, except for HS\,1157$+$3143), we ignore IGM absorption by setting the mean~free~path to  $\lambda_\mathrm{mfp}=\infty$.

We calculate \ion{H}{i} and \ion{He}{ii} ionization rates resulting from the quasar based on the \citet{Lusso2015} quasar template, assuming a power-law of slope $\alpha = -1.7$ beyond $912\,\mathrm{\AA}$.  
For simplicity, we assume that the spectral dependence of the ionization cross-sections of helium and hydrogen have a power-law  of form $\sigma_{\nu} \propto (\nu/\nu_0)^{-3}$, %
and take the cross-sections at the ionizing-edges $\sigma_0$ from \citet{Verner1996b}%
\footnote{The exact spectral dependence of the \ion{He}{ii} ionization cross-section is of low importance due to substantial uncertainty in the quasar extreme UV continuum and the \ion{H}{i} quasar ionizing rate is anyway lower than the UV background.}%
.
This leads to the \ion{H}{i} and \ion{He}{ii} quasar ionizing rates of the form 
\begin{equation}
 \Gamma_\mathrm{QSO}^\mathrm{ION} = \int_{\nu_o^\mathrm{ION}}^\infty \frac{ F_\nu \; \sigma_{\nu}^\mathrm{ION} }{ \mathit{h}_\mathrm{P} \, \nu } \: d\nu\; \approx \frac{ F_{\nu_o^\mathrm{ION}} \; \sigma_0^\mathrm{\,ION} }{\mathit{h_\mathrm{P}} \; (3 - \alpha)}
 \label{Eq:Q_QSO}
\end{equation}
in which $\mathit{h}_\mathrm{P}$ denotes Planck's constant and $\nu_0^\mathrm{ION}$ the frequency of the corresponding ionization edge.
Due to the different cross~sections and the chosen quasar spectral energy distribution, we find $\Gamma_\mathrm{QSO}^\mathrm{HI} \approx 42 \:  \Gamma_\mathrm{QSO}^\mathrm{HeII}$.
Evaluating $\GqsoHe(z)$ at the foreground quasar redshifts, therefore in exactly transverse direction, gives the $\Gamma_\mathrm{QSO,\:max}^\mathrm{HeII}$ values quoted in Table~\ref{Tab:Objects}.

The additional ionization by the quasar might also have an effect on the thermal structure of the IGM \citep{Bolton2009, Bolton2010, Bolton2012}. However, proper treatment of this \textit{thermal proximity effect} would require radiative transfer calculations \citep{Meiksin2010, Khrykin2017} which is beyond the scope of this study. Also, the thermal proximity effect for \ion{He}{ii} should be sub-dominant compared to the enhanced \ion{He}{ii} ionization \citep{Khrykin2016}.

In the following we calculate the regions of the background sightlines that are, depending on quasar age and obscuration, indeed illuminated by the foreground quasars.

\subsubsection{Quasar Obscuration}
\label{Sec:QuasarObscuration}

For the geometry of the foreground quasars radiation we assume a simple biconical emission model with half-opening angle $\alpha$ of the cones. Such an emission pattern is  suggested by the observations of local Seyfert galaxies \citep[e.g.][]{Tadhunter1989, Wilson1993} and the quasar unification scheme \citep{Antonucci1993, Urri1995, Netzer2015}.
The solid angle on the sky not illuminated by the quasar is then $\Oobsc = 4\,\pi \cos( \alpha )$. For simplicity, we usually state the obscured fraction of the sky (omitting the $4\,\pi$). For $\alpha=60^\circ$ half of the sky is illuminated ($\Oobsc=50\%$) and $\alpha = 90^\circ$ corresponds to isotropic emission ($\Oobsc=0\%$).
The orientation of the foreground quasar's emission bicone with respect to the background sightline is described by two angles $(\theta, \phi)$. Here, $\theta$ denotes the angle between the quasars polar axis and the line of sight (\textit{inclination}) where $\theta=0^\circ$ describes the case in which the polar axis points directly towards Earth. The apparent direction on the sky, as seen from Earth, in which the quasars polar axis is tilted (\textit{position angle}) is denoted with $\phi$.  A bicone pointing
towards the background sightline corresponds to $\phi=0^\circ$, $\phi=90^\circ$ perpendicular to it and $\phi = 180^\circ$ away from it.

For a given point on the background sightline the angle between the foreground quasar polar axis and a ray from the foreground quasar towards this point is
\begin{equation}
\beta            = \frac{ \: \arccos( R_\perp \sin(\theta) \cos(\phi) - R_\parallel \cos(\theta) \; ) }{ \sqrt{ R_\parallel^2 + R_\perp^2 } } 
\end{equation}
in which $R_\perp$ denotes the comoving separation between the foreground quasar and background sightline and $R_\parallel$ the comoving distance along the background sightline, measured from the point of closest approach towards the background quasar. All locations for which $\beta < \alpha$ or $\beta > 2\,\pi - \alpha$ are illuminated. All other positions do not receive any quasar radiation. Instead the foreground quasar appears as an obscured Type\,II from these vantage points.

Within our model, the quasar half-opening angle $\alpha$ or equivalently the fraction of the sky which is obscured, $\Oobsc$,  is chosen explicitly while the quasar orientation $(\theta, \phi)$ is randomly drawn. The foreground quasars in our sample appear as unobscured Type\,I from Earth. This constrains the orientation to $\theta < \alpha$. We achieve this by drawing $\phi$ from a flat distribution between $0 < \phi < 2\,\pi$ and $\cos(\theta)$ from a flat distribution between $1 > \cos( \theta ) > \cos( \alpha )$.

\subsubsection{Finite Quasar Age}
\label{Sec:QuasarLifetime}

A key element for our sensitivity to quasar age is the fact that the background sightline probes the foreground
quasars emission at earlier times than the light we directly receive from the quasar \citep[see e.g.][]{Adelberger2004, Kirkman2008, Furlanetto2011, Schmidt2017}.
This arises because of the geometric path length differences  between the longer path from the foreground quasar to a location along the background sightline, and from there to the observer (as probed by the background sightline), compared to the direct path
from the foreground quasar to Earth.
The relevant quantities to compute this path length difference are the distance (from Earth) to a location along the background
sightline at redshift $z$, and the distance from this point to the foreground quasar%
\footnote{The comoving distance between a location on the background sightline and the foreground quasar for an angular sightline separation of $\Delta{}\theta$ can be computed via $\sqrt{ R_\parallel^2 + R_\perp^2 } \equiv r( \, z, z_\mathrm{QSO}, \Delta{}\theta \, )= \sqrt{ \, r(z)^2 + r(z_\mathrm{QSO})^2 - 2 \, r(z) \, r(z_\mathrm{QSO}) \cos( \Delta{}\theta ) \, } $ \citep[e.g.][]{Liske2001} }%
. When measuring both distances in comoving units, their sum can be converted to a redshift $z_\mathrm{em}$ and corresponding lookback time $t_\mathrm{em}$ at which the ionizing radiation from the foreground quasar had to be emitted.

\begin{figure}
 \centering
 \includegraphics[width=\linewidth]{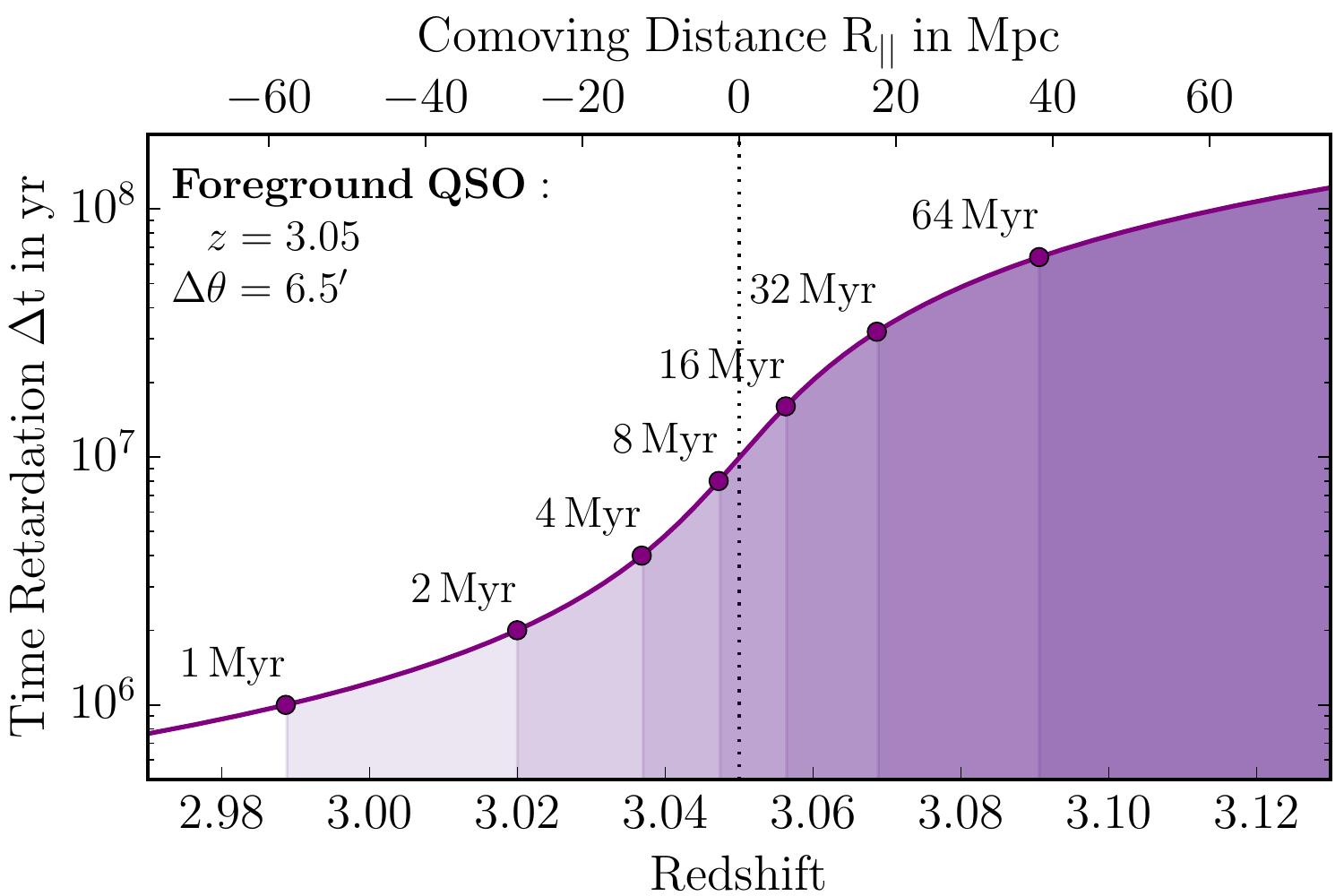}
 \caption{Visualization of the time retardation $\Delta{}t$ along the Q\,0302$-$003 sightline with respect to the $z=3.05$ foreground quasar at angular separation of $\Delta\theta = 6.5'$. For a given age of the foreground quasar $\tTO$, only the part of the background sightline for which $\Delta{}t < \tTO$ appears illuminated.
 }
 \label{Fig:TimeRetardation}
\end{figure}

The lookback time at emission can be compared to the lookback time corresponding to the redshift of the foreground quasar $z_\mathrm{QSO}$. 
The difference is the additional time $\Delta{}t(z)$ it takes to first reach a certain point on the background sightline. This \textit{time retardation} depends on the redshift of the point in question and of course quasar redshift and sightline separation. For points at redshifts lower than the foreground quasars ($z<z_\mathrm{QSO}$) the time difference $\Delta{}t(z)$ is relatively small. For $z=z_\mathrm{QSO}$ it is exactly the transverse light crossing time $\Delta{}t = R_\perp \, \mathrm{c}^{-1}$ with the transverse separation $R_\perp$ now measured in proper length and $\mathrm{c}$ denoting the speed of light. For positions at higher redshift than the foreground quasars, therefore behind it, $\Delta{}t(z)$ quickly increases. See Figure~\ref{Fig:TimeRetardation} for a visualization.

Whether or not a given location along the background sightline is illuminated  now depends on the age of the foreground quasar, since there had to be enough time for its ionizing radiation to arrive at a given location (Figure~\ref{Fig:TimeRetardation}).
For the quasar lightcurve we assume a simple lightbulb model in which the quasar turns on and shines with constant luminosity for its entire lifetime $t_{\rm Q}$. In this case, the age of a quasar $\tTO$ is well defined and represents the time between turning on and emission of the photons that arrive at Earth today. For a discussion about more complicated quasar lightcurves see \S~\ref{Sec:Non-Lightbulb}. 
Points on the background sightline for which $\Delta{}t(z) < \tTO$ appear for an observer on Earth illuminated by the quasar.
Since $\Delta{}t(z)$ monotonically increases with $z$ (see Figure~\ref{Fig:TimeRetardation}), all points at redshifts higher than the dividing line where $\Delta{}t(z) = \tTO$ appear not yet illuminated since there was not enough time for the photons to reach these locations.

\subsection{Example of the Simulated Data}

\begin{figure*}
 \centering
 \includegraphics[width=.8\linewidth]{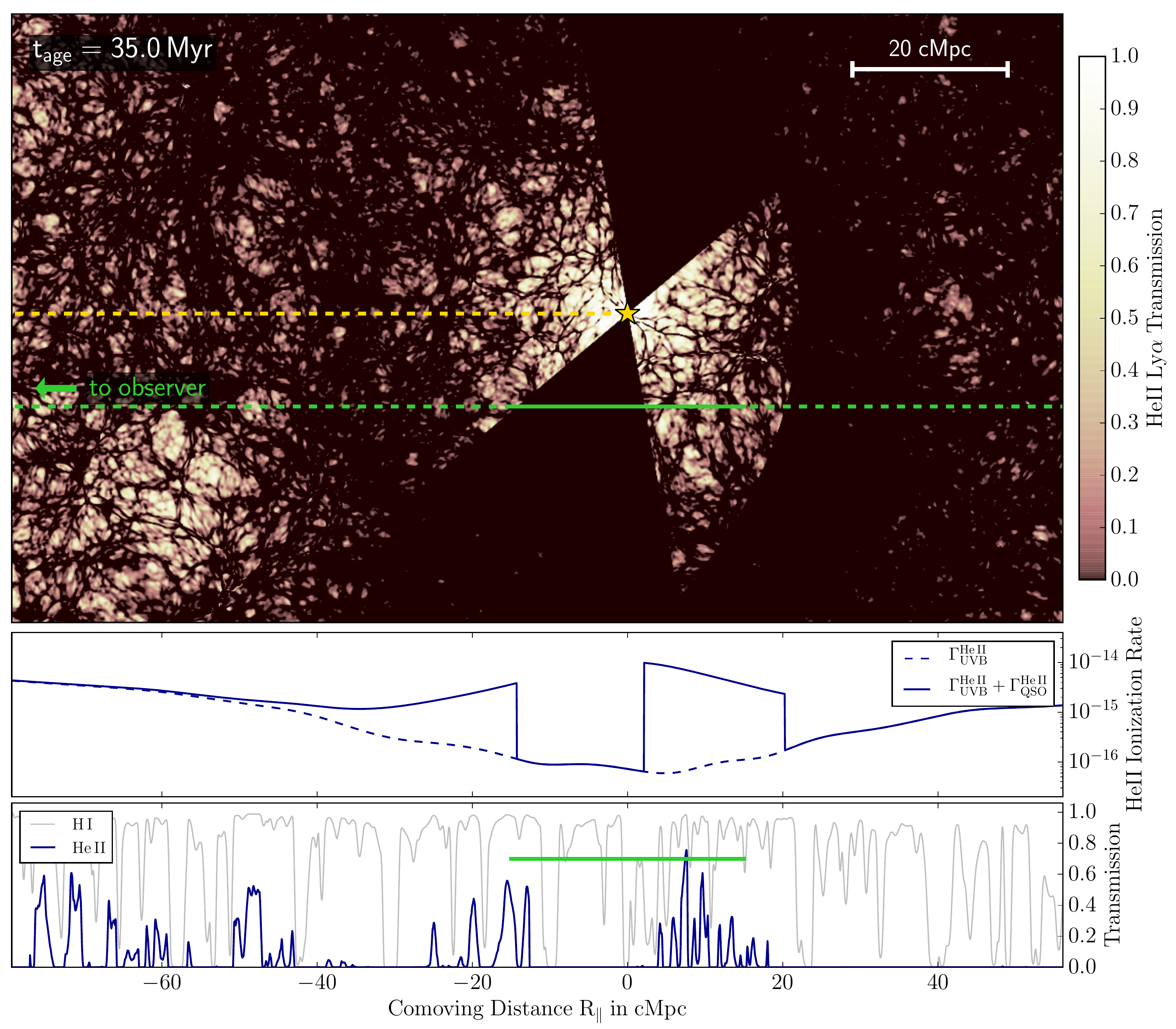}
 \caption{
  Illustration of our model, showing quasar obscuration, finite quasar age and \ion{He}{ii} UV background fluctuations. The top panel displays for a slice through the simulation the \ion{He}{ii} transmission in realspace, clearly showing the bi-conical emission of the quasar and the parabolic shaped region that can be reached for the given quasar age. The middle panel indicates the resulting \ion{He}{ii} ionization rate along the green marked background sightline. The bottom panel shows the computed hydrogen and helium transmission spectra, as observed in redshift space. A strong \ion{He}{ii} transmission enhancements is visible in regions that are illuminated by the quasar. The solid green bar marks the the $\pm15\cMpc$ we defined as \textit{proximity region} in \citet{Schmidt2017}.
 }
 \label{Fig:2D_Example}
\end{figure*}

In Figure~\ref{Fig:2D_Example} we visualize one of our photoionization models. Sightline geometry and quasar luminosity are matched to the foreground quasar at $z=3.05$ along the Q\,0302$-$003 sightline \citep{Jakobsen2003}. The top panel shows the computed \ion{He}{ii} $\lya$ transmission in a slice through the simulation box as it would appear for an observer on Earth%
\footnote{Not including peculiar velocities / redshift space distortions}%
. The quasar is placed in a $10^{12}\Msun$ halo and emits in a biconical pattern with $\alpha=60^\circ$ and therefore illuminates half of the sky. It is tilted by $\theta=20^\circ$ against the line of sight towards the observer (yellow). The assumed finite quasar age of $\tTO=35\Myr$ limits the extend of the ionized area towards the right. The positions for which the quasar emission had sufficient time to reach them lie in a parabolic shaped region with the quasar at the focal point. This parabola expands with increasing quasar age.

The middle panel of Figure~\ref{Fig:2D_Example} shows the \ion{He}{ii} ionization rate along the background sightline (green) separated from the quasar by $R_\perp\approx12\cMpc$. Clearly visible is the effect of quasar obscuration ($-17\cMpc < R_\parallel < 2\cMpc$) and finite quasar age ($R_\parallel>20\cMpc$).

The bottom panel shows synthetic \ion{H}{i} and \ion{He}{ii} transmission spectra along the sightline. No transverse proximity effect is visible for hydrogen but a clear enhancement in \ion{He}{ii} transmission is visible in regions that are illuminated by the foreground quasar, e.g. around $R_\parallel \approx -20\cMpc$ and $R_\parallel \approx 10\cMpc$. One can see that this \ion{He}{ii} transmission is highly modulated by the cosmic density structure, as traced by the \ion{H}{i} Ly$\alpha$ absorption.  Whenever there is a substantial \ion{H}{i} absorber, we observe saturated \ion{He}{ii} absorption.
On the other hand, substantial \ion{He}{ii} transmission is not necessarily associated with the presence of our bright foreground quasar. It can also be caused by the fluctuating \ion{He}{ii} UV background as can be seen in the top panel of Figure~\ref{Fig:2D_Example} in particular in the lower left corners of the transmission slice. For $R_\parallel<-40\cMpc$ along the sightline through the box (lower panel of Figure~\ref{Fig:2D_Example}) these fluctuations result in \ion{He}{ii} transmission nearly comparable to the values in the transverse proximity zone, despite having no explicit foreground quasar there. Such situations are indeed consistent with observations. In \citet{Schmidt2017} we showed several strong transmission spikes without a clear association to foreground quasars and our test in \S~\ref{Sec:HeII_UVB} showed that large fluctuations in the \ion{He}{ii} UV background are actually required to match the observed data.

The region we defined as the \textit{proximity region} and used to measure the \ion{He}{ii} transmission enhancement in \citet{Schmidt2017}, $\pm15\cMpc$ around the foreground quasar position, is indicated in Figure~\ref{Fig:2D_Example} as a solid green bar. For the quasar orientation shown in Figure~\ref{Fig:2D_Example}, a substantial part of this region is not illuminated by the quasar and shows no transmission enhancement while at the same time some extra transmission falls outside the chosen window. However, the signature of the transverse proximity effect is highly stochastic and depends not only on sightline geometry, quasar age, and obscured sky area but in particular on the random orientation of the quasar, cosmic density structure and UV background fluctuations. It is therefore important to investigate the statistical properties of the expected transverse proximity signal which we
address in the next section.

\subsection{Average Transmission Profiles}
\label{Sec:Model_Averages}

\begin{figure*}
 \centering
 \includegraphics[width=\linewidth]{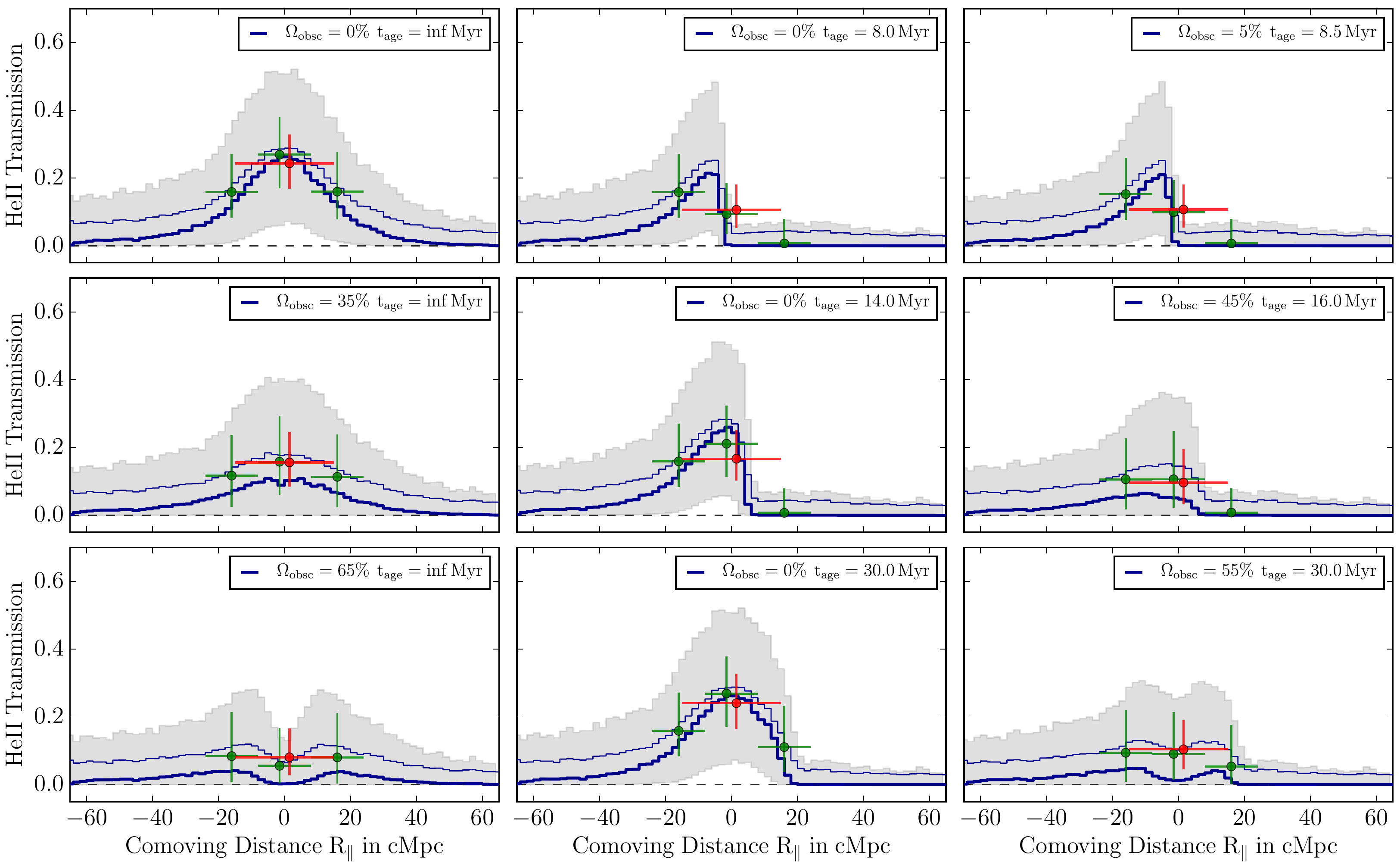}
 \caption{
  Visualization of our models for different quasar properties. Sightline geometry and quasar luminosity are matched to the Q\,0302-003 $z=3.05$ object. The thick blue line shows the median \ion{He}{ii} transmission in $2\cMpc$ bins, the gray shaded area the 16th\,--\,84th percentile scatter. We also show the mean transmission as thin blue line. Due to the large non-Gaussianities in the distributions these can be vastly different from the median. The colored points show the median and expected scatter of the \ion{He}{ii} transmission averaged over $16\cMpc$ (green) and $30\cMpc$ (red) wide windows. Symbols are slightly displaced for clarity.
  Models in the first column show the effect of obscuration, models in the second column lifetime effects. Models in the last column are selected to give the same average transmission over the $\pm15\cMpc$ window but have different signal shapes that in principle could be distinguished using transmission measurement in three bins.
 }
 \label{Fig:Model_Averages}
\end{figure*}

We illustrate the average \ion{He}{ii} transmission profile and the associated scatter for models with different quasar
properties in Figure~\ref{Fig:Model_Averages}. 
For each of these models we compute a large number
of skewers sampling our \ion{He}{ii} UV background model and IGM density fluctuations, with a fixed $\Oobsc$ (i.e. $\alpha$) and $\tTO$ but randomly drawn quasar orientation ($\theta$, $\phi$). 
The quasar luminosity and sightline geometry are again chosen to match the Q\,0302$-$003 $z=3.05$ foreground quasar.
The dark blue lines in Figure~\ref{Fig:Model_Averages} represent the averages (mean and median) of 2000 skewers, each binned to $2\cMpc$ bins, approximately the typical pixel size of \ion{He}{ii} spectra. The gray shaded region represents the scatter (16th to 84th percentile region) within the set of 2000 skewers. Observational effects like photon counting noise are not included here. Instead, only the variance within the models is shown. 
This illustrates the extreme stochasticity of the \ion{He}{ii} transverse proximity effect and the concomitant challenge
of interpreting single absorption spectra.

However, note that the transmission values in the small $2\cMpc$ bins are highly correlated and the transmission distribution highly non-Gaussian. To better illustrate the expected variance we show a synthetic measurement of the transmission averaged over our chosen window of $\pm15\cMpc$. The red point shows the median value for this measurement. The horizontal bar indicates the size of the region while the vertical bar indicates the expected scatter in this measurement derived from the 16th and 84th percentile of the distribution. 
In addition, we show measurements in three consecutive $16\cMpc$ wide bins (green points) that allow one to to better
capture the shape of the signal. 

The upper left panel of Figure~\ref{Fig:Model_Averages} shows a model for which the quasar emission is isotropic and the quasar
age infinite. The scatter therefore arises from density and \ion{He}{ii} UV background fluctuations alone. 
The other panels in the left column show models which also have infinite age, but with the quasar emission restricted
to $65\%$ and $35\%$ of the sky. This clearly reduces the amplitude of the transverse proximity effect signal, and for the
$35\%$ model even results in a dip in the average transmission at $R_\parallel=0\cMpc$. Here, the quasar emission is so highly beamed that it may hit the background sightline in front and behind the foreground quasar and causes additional \ion{He}{ii} transmission there ($R_\parallel \approx \pm 15\cMpc$),  but since it is constrained to shine towards it Earth basically cannot illuminates the background sightline at $R_\parallel = 0\cMpc$.

The second column shows models with isotropic emission but varying quasar age between $8$ and $30\Myr$. As described above, only points on the background sightline for which the time retardation is shorter than the quasar age ($\Delta{}t(z)<\tTO$) can be reached by the quasar radiation and therefore show enhanced transmission.
These point all lie to the left (lower redshifts, lower $R_\parallel$) of where $\tTO=\Delta{}t$. To the right of this, one only observes transmission caused by the \ion{He}{ii} UV background. With increasing quasar age, this cut-off moves to the right (higher redshift, higher $R_\parallel$). The position of the cut-off is of course also influenced by the separation between foreground quasar and background sightline.

The right column in Figure~\ref{Fig:Model_Averages} shows models with different combinations of quasar age and obscuration. The three sets of model parameters ($\Oobsc$ and $\tTO$) are selected to give approximately the same transmission enhancement in the $\pm15\cMpc$ window (red measurements). Since the quasar lifetime has a very asymmetric effect on the background sightline, it is, at least in principle, possible to break this degeneracy by measuring the transmission enhancement in multiple bins (green points). However, the large estimated scatter in the measurement (again, this includes only model stochasticity, no measurement uncertainties) sets limits on the confidence with which these models can be distinguished.

In general, one can deduce from Figure~\ref{Fig:Model_Averages} that distinguishing different models at very high significance will probably not be possible. The expected variance in the \ion{He}{ii} transverse proximity effect is simply too high in single spectra. However, it should be possible to rule out some extreme cases and broadly distinguish between scenarios. This however requires a sophisticated statistical analysis and calls for a fully Bayesian approach that can naturally deal with non-Gaussian distributions, strong degeneracies, and weekly constrained parameters, which is our task in the next section.

\section{Comparison to Data and Inference of Parameters}
\label{Sec:4}

Our aim is to infer individual quasar ages and obscuration properties for the six foreground quasars with the highest estimated \ion{He}{ii} photoionization rate at the background sightline. In several cases, there might not be be a single definitive answer to this. We however intend to calculate, in a fully Bayesian way, the joint probabilities for a wide range of $\Oobsc$\:--\:$\tTO$ combinations which then hints towards certain regions in the parameter space or rules out others.

To this end, for each foreground quasar we compute a grid of models that covers the parameter space from $\Oobsc = 5\%$ to $\Oobsc = 95\%$ and quasar ages from $5\Myr$ to $46\Myr$. 
For simplicity, we decided to sample the parameter space with a rectangular model grid of size $10 \times 12$ for ($\Oobsc$, $\tTO$) and avoid any interpolating between models but instead just evaluate the likelihood at the points of the model grid. Since our constraints will be broad anyway, this is not a substantial disadvantage.
To properly capture the stochasticity in the \ion{He}{ii} transverse proximity effect and to adequately map the distribution of the expected \ion{He}{ii} transmissions we calculate 5000 skewers per model with randomly drawn quasar orientation, and sampling of the UV background and cosmic density field along the different skewers. This then allows us to infer the probability of each model given the observed data.

\subsection{Likelihood Computation}
\label{Sec:Likelihood}

To simplify the explanation of the likelihood calculation and make the it easier to understand for the reader, we adopt for this part the mean \ion{He}{ii} transmission statistic $\F$. However, for the actual analysis we use the transmission enhancement statistic $\xiTPE$ (see Equation~\ref{Eq:xiTPE}). The necessary modifications to the likelihood computation are straight forward an described later in \S~\ref{Sec:TransmissionEnhancementStatistic}.

Our measurement in the spectra are the photon counts $\CT_i$ in pixels $i=1 \dots N$. Additional information computed during the data reduction are the sensitivity function $\SE_i$, the exposure time $\EX_i$\footnote{The exposure time varies from pixel-to-pixel, in particular due to grid wires in front of the COS FUV detector}, the total number of expected background counts $\BG_i$ and a fit for the quasar continuum $\CO_i$. For details see \citet{Worseck2016}.
These information are sufficient to translate the measured counts into transmission values. 
However, the Poisson nature of the count distribution requires a forward modeling to calculate proper uncertainties. Often, the detector received only a handful of counts per pixel, but in regions of saturated absorption this can be as low as one or zero source counts. Clearly, assuming Gaussian errors, described by mean and standard deviation, is not appropriate for our case. Instead, we have to propagate full Poisson errors.

Our model parameters are quasar age $\tTO$ and obscured sky fraction $\Oobsc$. We therefore have to compute the following likelihood:
\begin{equation}
\mathcal{L} = p( \CT_{i=1 \dots N} | \SE_i, \EX_i, \CO_i, \BG_i, \tTO, \Oobsc )
\label{Eq:1}.
\end{equation}
Instead of applying the complete forward modeling directly to our skewers, we separate the measurement process from the IGM physics.
The first part only deals with the noisy detection process and therefore measurement uncertainties, the second part represents the physics of the \ion{He}{ii} transverse proximity effect and captures the associated stochasticity. 

To make this separation, we introduce the intrinsic, noise-free average transmission $\F$, measured over a given bin, as an intermediate quantity (\textit{observable}). 
In practice, we extract from the models the \ion{He}{ii} transmission averaged over the region $\pm15\cMpc$ around the foreground quasar position 
\begin{equation}
  \F = \langle \F_{|R_\parallel|<15\cMpc}\rangle
\end{equation}
and then assume this average value for the forward modeling of the photon-counting noise%
\footnote{This simplification has only minimal impact on the precision of our noise estimate. Also, the intrinsic variance in the transverse proximity effect anyway dominates over the photon-counting noise.}.
The separation of IGM physics and measurement process not only saves a large amount of computation time but is also intuitive. It can be formally written as
\begin{equation}
p( \CT_i | \tTO, \Oobsc ) = \int \; p( \CT_i | \F ) \; p( \F | \tTO, \Oobsc ) \; d\F
\label{Eq:2} \: .
\end{equation}
The first term in the integral is the Poisson probability of measuring the counts $\CT_i$ given an intrinsic transmission $\F$ within the bin:
\begin{equation}
P_{\CML_i}(\CT_i) = \frac{ {\CML_i}^{\,\CT_i} }{\CT_i !} \: \mathrm{e}^{-\CML_i}
\label{Eq:3}
\end{equation}
with the definition of the most-likely photon count
\begin{equation}
\CML_i = \F_i \cdot \CO_i \cdot \SE_i \cdot \EX_i + \BG_i
\label{Eq:3a}
\end{equation}
which combines continuum estimate, sensitivity, exposure time and total background counts for each individual pixel. 
These values are derived within the data reduction process described in \citet{Worseck2016}.

At this point it is convenient to combine all pixels within the selected bin, denoted with $\CT = \{ \CT_i \}$, to the joint probability%
\begin{equation}
p( \CT | \F ) = \prod_i  \; p( \CT_i | \F )
\label{Eq:4} \:.
\end{equation}
This operation is permitted since the photon-counting noise in the individual pixels is uncorrelated and $\F$ represents the transmission averaged  over the bin and is therefore a constant.
The in Equation~\ref{Eq:4} computed probability reflects the combined measurements of many pixels and the resulting probability distribution is therefore more Gaussian than the Poisson distributions of the individual pixels. We make use of this for the noise estimate in \S~\ref{Sec:HeII_UVB}. and avoid propagating single pixel Poisson noise for that case. \label{Sec:NoiseMeasurement}

The second term in Equation~\ref{Eq:2}, $p( \F | \tTO, \Oobsc )$, represents the expected \ion{He}{ii} transmission along a sightline given our model parameter $\tTO$ and $\Oobsc$. Since quasar orientation, \ion{He}{ii}~UV background and cosmic density structure are stochastic, this term is not a single value but as shown in Figure~\ref{Fig:Model_Averages} a broad distribution which we sample with 5000 skewers per model. 
To overcome the discrete sampling of  $p( \F | \Oobsc, \tTO )$ caused by the finite number of skewers, we apply a kernel density estimate (KDE) with Scott's~rule for the kernel width to approximate the distribution. The KDE makes $p( \F_i | \tTO, \Oobsc )$ a smooth and continuous function and at the same time ensures that the probability is nowhere exactly zero, which would lead to numerical problems. 

After estimating the distribution of our observable in this way, we can finally compute the integral in Equation~\ref{Eq:2}. This is done via a discrete Monte~Carlo approach by randomly sampling the KDE with $10^5$ points and evaluating the Poisson distribution (Equation~\ref{Eq:3} and \ref{Eq:4}) for each sample. Averaging these samples yields the desired likelihood in Equation~\ref{Eq:1}.

\subsubsection{Transmission Enhancement Statistic $\xiTPE$}
\label{Sec:TransmissionEnhancementStatistic}

While the above description illustrates our approach using the intrinsic transmission $\F$ as the main observable,
we prefer to  use the flux enhancement statistic $\xiTPE$,  since it is to first order independent of the \ion{He}{ii} mean transmission
and better isolates the effect of the foreground quasar (see Figure~\ref{Fig:Test_UVB}). As already mentioned in \S~\ref{Sec:Sample}, $\xiTPE$ is similar to the statistic used in  \citet{Schmidt2017}, and is defined as the difference between the transmission in the proximity region and in a wider background region:
\begin{equation}
\xiTPE = \langle \F_{|R_\parallel|<15\cMpc}\rangle - \langle \F_{15\cMpc<|R_\parallel|<65\cMpc}\rangle
\label{Eq:xi2}.
\end{equation}
Calculating the likelihood using $\xiTPE$ is essentially identical to the approach outlined above, by simply
replacing $\F$ by $\xiTPE$. Complications only arise in including the measurement
uncertainty in the background transmission (outside the $\pm15\cMpc$ proximity region, second term in Equation~\ref{Eq:xi2}). 
The background transmission is calculated over $3\times$ the pathlength of the the $\pm15\cMpc$ proximity region and the photon counting noise is therefore far less important. Despite this, we propagate the associated uncertainty in a Bayesian way into our analysis, but refer the reader to the Appendix for the exact mathematical description. 
The remaining part of the likelihood calculation is completely analogous and just requires a computation of $\xiTPE$ in the simulated skewers instead of $\F$.

\subsubsection{Three-Bin Statistic}

As illustrated in Figure~\ref{Fig:Model_Averages} and discussed in \S~\ref{Sec:Model_Averages}, degeneracies may arise between the parameters $\tTO$ and $\Oobsc$, in particular if we extract only one transmission measurement from the spectra. However, this degeneracy can to some degree be broken by measuring the \ion{He}{ii} transmission in several consecutive bins along the background sightline. The last column of Figure~\ref{Fig:Model_Averages} shows three models with different combinations of $\tTO$ and $\Oobsc$ that result in a nearly identical \ion{He}{ii} transmission measured over $\pm15\cMpc$, but due to the asymmetric effect of quasar age show different transmission levels in the three $16\cMpc$ wide bins.
We therefore try to use this additional information about the signal shape to better disentangle $\tTO$ and $\Oobsc$ effects.
However, this significantly complicates our statistical method. Instead of one transmission in the $\pm15\cMpc$ region we have to deal with multiple (e.g. three) transmission measurements and our observable $F$ becomes a multi-dimensional quantity $F_k$. Calculating a full Bayesian likelihood for such a multi-bin measurements including all correlations is extremely challenging in the context of our study.

Based on our analysis, we conclude that the transmission values in multiple bins are highly correlated. It is therefore not possible to separate the likelihood computation into three one-dimensional problems. Due to the non-Gaussian nature of \ion{He}{ii} transmission (illustrated in Figure~\ref{Fig:Model_Averages}) it is also not possible to assume that a multivariate Gaussian distribution describes this multivariate process.  Indeed, this is already the reason we could not condense our models to mean and standard deviation in the one-dimensional case.
The only possible approach in our view is (again) a full description of the multivariate probability distribution. 

Mathematically, this is simple. Nothing in the procedure outlined above for computing the likelihood (\S~\ref{Sec:Likelihood}) assumes the transmission $F$ to be a one-dimensional quantity. In principle, the approach can be extended to arbitrary dimensionality.
However, the computational effort for this brute-force method increases dramatically with increasing dimensionality. 

The required increase in the number of Monte Carlo evaluations from $10^5$ to $10^7$ for computing the integral in Equation~\ref{Eq:2}  is merely an inconvenience.
The ultimate limitation however poses the immense number of mock skewers required to properly sample the multivariate probability distribution in high dimension. For a measurement in three bins we found 5000 simulated skewers per model to be sufficient to adequately map the probability density. For more dimensions, the required number of skewers quickly increases dramatically.
The number of three bins therefore poses the practical maximum. Also, when using more but smaller bins, the stochasticity of the measurements increases and it becomes increasingly important to have a correct model for the small-scale fluctuations in the \ion{He}{ii} UV background. 

In practice, we implement the three-bin statistic in complete analogy with the single bin measurement, and similarly use $\xiTPE$ rather than the raw \ion{He}{ii} transmission $F$. For the three bins, the transmission is measured in three consecutive $16\cMpc$ wide windows between $-24\cMpc < R_\parallel < 24\cMpc$ and the background for all three bins in the region $24\cMpc<|R_\parallel|<65\cMpc$.

\subsection{Priors}

We impose uninformative  flat priors on $\Oobsc$ and $\tTO$. 
A strong prior by itself however is the extent of the parameter grid.
For $\Oobsc$ we explore the full possible range from nearly isotropic emission ($\Oobsc=5\%$) to almost complete obscuration ($\Oobsc=95\%$) in steps of $10\%$. For $\tTO$ we limit our analysis to possible quasar ages between $5\Myr$ and $46\Myr$ since the sightline geometries for the six foreground quasars allow only very limited sensitivity to timescales outside this range.

\section{Results}
\label{Sec:Results}

\begin{figure*}
\begin{center}
\includegraphics[width=\linewidth]{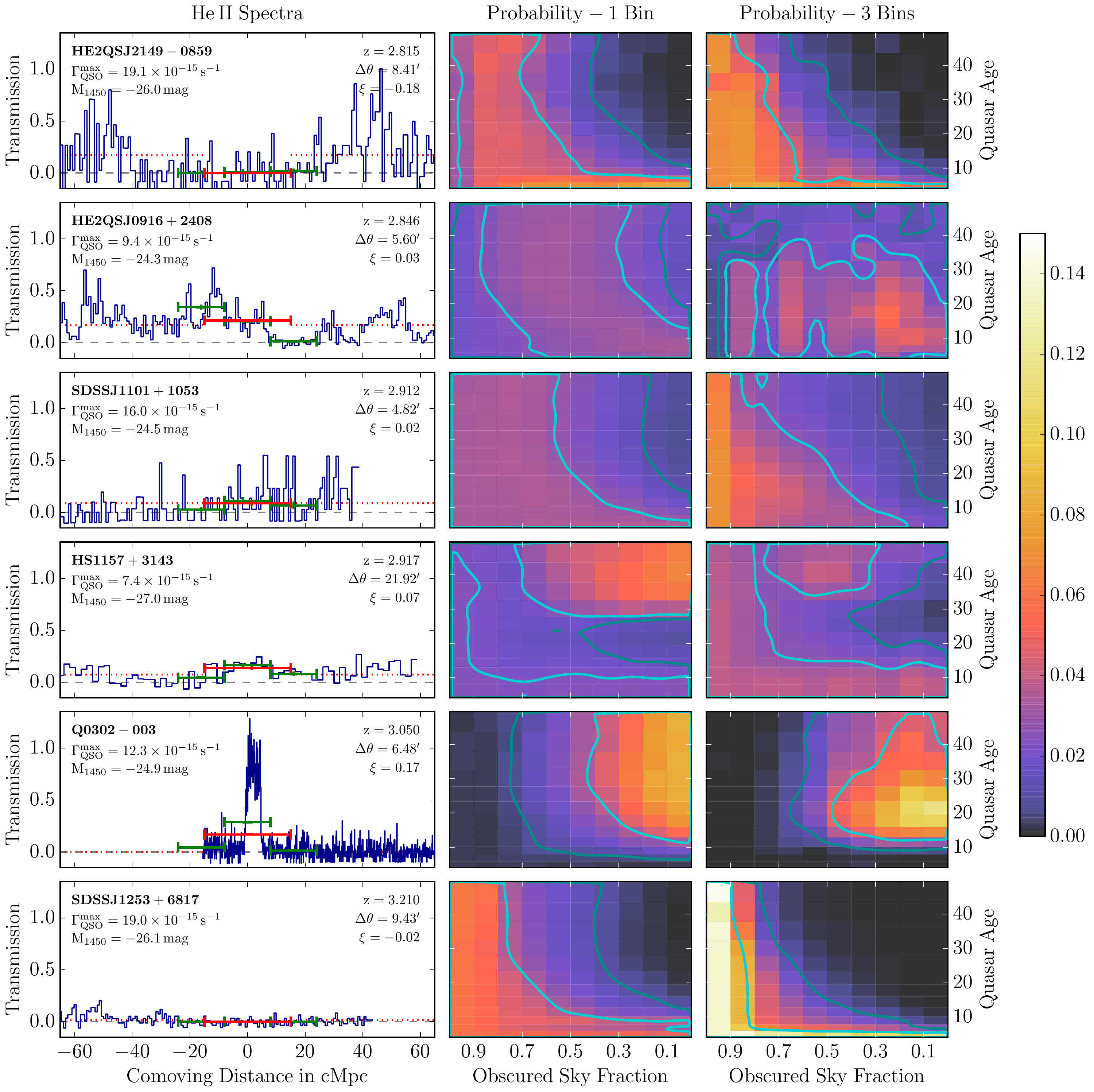}
\end{center}
\caption{Joint posterior probabilities for age and obscured sky fraction of the six quasars analyzed in this study. The left column shows the FUV spectra of the \ion{He}{ii} background sightline. The red bars marks the measured \ion{He}{ii} transmissions in the $30\cMpc$ wide bins, the dotted red lines indicate the background transmissions. The green bars show the measured transmission in the three $16\cMpc$ wide bins. Some key information about the foreground quasars are given as well.
The middle and right columns give the inferred posterior probabilities for quasar age and obscured sky fraction. Bright and dark cyan contours indicate $1$ and $2\,\sigma$ regions. The probabilities in the central column are derived using the transmission enhancement in a single $\pm15\cMpc$ wide window. The ones in the right column are based on the transmission enhancement in three $16\cMpc$ wide bins. Both statistics give consistent results but the three-bin statistic has improved sensitivity to quasar age.
For the three foreground quasars with the highest ionization rate, our analysis prefers scenarios in which the quasars are very young ($<10\Myr$) or highly obscured ($\Oobsc>70\%$), which in both cases would prevent ionizing radiation from reaching the background sightline.
For the foreground quasar associated with the large transmission spike in the Q\,0302$-$003 sightline, we find low obscuration ($\Oobsc<40\%$) and an age above $15\Myr$ with a peak probability in the three-bin statistic around $22\Myr$.
For the other two quasars we derive only very weak constraints or even bimodal distributions.  
}
\label{Fig:Key_Plot}
\end{figure*}

For the six foreground quasars with the highest \ion{He}{ii} ionization rate from \citet{Schmidt2017} we have modeled the \ion{He}{ii} transmission along the background sightline and derived joint constraints on quasar age $\tTO$ and obscured sky fraction $\Oobsc$. The results are shown in Figure~\ref{Fig:Key_Plot}. The left column shows $130\cMpc$ long sections of the observed \ion{He}{ii} transmission spectra around the position of the six analyzed foreground quasars. 
The red horizontal bars indicate the average transmission in the $\pm15\cMpc$ window $\langle \F_{|R_\parallel|<15\cMpc}\rangle$ and the dotted lines the background transmission $\langle \F_{15\cMpc<|R_\parallel|<65\cMpc}\rangle$. The transmission enhancement $\xiTPE$ is the difference of these averages as given in Equation~\ref{Eq:xi2}.
Green horizontal bars indicate in full analogy the transmission measured in the three $16\cMpc$ wide bins.
Measurement error are usually $<3\%$ and therefore not visible in the plot.

Middle and right column of Figure~\ref{Fig:Key_Plot} show the joint posterior probabilities for quasar age and obscured sky fraction. The results in the central column are derived from the transmission enhancement measured in the single $30\cMpc$ wide bin, the ones in the right columns from the transmission enhancement statistic in three consecutive $16\cMpc$ wide bins. The $1\,\sigma$ and $2\,\sigma$ contours (enclosing 68\% and 95\% total probability)  are shown as bright and dark cyan lines, determined by smoothing the pixelated likelihood surface. 

In general, the results from the single-bin statistic and the three-bin statistic are in agreement which is highly encouraging given the substantial differences between the statistics and the 60\% longer pathlength used in the three-bin statistic.
Consistent with expectation, the three-bin contours are typically slightly better constrained than the contours derived using the single bin statistic.
Note that in basically all cases, the contours are not closed. We therefore obtain just limits on the parameters, in particular the quasar age%
\footnote{The obscured sky fraction is naturally constrained between 0\% and 100\%. Contours not closed in $\Oobsc$ therefore have slightly different quality than contours open in $\tTO$ direction.}%
. This was expected given the high level of fluctuations in our transverse proximity models illustrated in Figure~\ref{Fig:Model_Averages}.  
Since the likelihood distributions are not localized, the adopted priors do have a substantial effect on the posterior probabilities. 

Based on our analysis of the six quasars shown in Figure~\ref{Fig:Key_Plot},  a very heterogeneous picture emerges. 
For the objects along the HE2QS\,J2149$-$0859 (top) and SDSS\,J1253$+$6817 (bottom) sightlines, our analysis rules out combinations of long lifetime ($>15\Myr$) and substantial illumination ($>50\%$) and indicates that these objects are either very young ($\tTO<8\Myr$) or highly obscured ($\Oobsc>70\%$). Both cases have in common that no ionizing radiation from the foreground quasar reaches the background sightline and no excess \ion{He}{ii} transmission is observed. Given that
we do not see evidence for a transverse proximity effect at the background sightline, it is not
possible to discriminate between these degenerate cases.
For the SDSS\,J1101$+$1053 sightline (third row) the picture is similar but less constrained. Our $1\,\sigma$ contour encloses the full lower-left corner (low $\tTO$, high $\Oobsc$) of our parameter space and our analysis only securely rules out the extreme case of $\Oobsc < 20\%$ and $\tTO > 20\Myr$.

We find a totally different result for the $z=3.05$ quasar \citep{Jakobsen2003} along the Q\,0302$-$003 sightline. This is the only foreground quasar associated with a strong \ion{He}{ii} transmission spike and therefore our analysis prefers scenarios in which a large amount of ionizing radiation reaches the background sightline, therefore low obscuration ($\Oobsc<40\%$) and quasar ages longer than $15\Myr$ (fifth row in Figure~\ref{Fig:Key_Plot}).
The Q\,0302$-$003 sightline also shows the strongest difference between single and three-bin statistic.
The posterior probability for the single-bin statistic is constant for quasar ages above $30\Myr$, because for lifetimes
this long, the quasar radiation would modify the transmission at $R_{\parallel} > 15\cMpc$ (see Figure~\ref{Fig:TimeRetardation}), outside the window used for the single-bin statistic.  
The rightmost of the three small bins however extends to higher comoving distance and is therefore sensitive to longer quasar ages. For the three-bin statistic, we thus find that the posterior probability decreases towards high quasar ages and the $1\,\sigma$ contour is almost closed with a peak around $22\Myr$. The analysis therefore associates the right cutoff of the transmission peak ($R_\parallel \gtrsim 6\cMpc$) with some probability to a finite age of the quasar. Quasar ages substantially longer than $30\Myr$ become less likely, but are however not ruled out at high significance.

For the quasar close to the HE2QS\,J0916$+$2408 sightline, the posterior probability derived from the single bin statistic is rather flat (second row in Figure~\ref{Fig:Key_Plot}). The shape of the $1\,\sigma$ contour does look different for the three-bin statistic, but the actual probabilities are not that different. In both cases, a large fraction of the probed parameter space is allowed. The three-bin statistic slightly disfavors quasar ages longer than $35\Myr$, probably related to the very low \ion{He}{ii} transmission around $R_\parallel\approx16\cMpc$.

For the HS\,1157$+$3143 sightline (forth row) we measure a generally low \ion{He}{ii} transmission with a slight increase in the $|R_\parallel|<15\cMpc$ region.
Our analysis does not clearly indicate whether this small enhancement is caused by the foreground quasar or by a UV background fluctuation.
This is clear from our posterior probability distributions, which  are clearly bi-modal.
If the extra transmission is due to fluctuations in the UV background, the quasar should not illuminate background sightline, indicated by the $1\,\sigma$ contour encompassing high obscured sky area ($>80\%$) and young quasar age ($<15\Myr$).
If the transmission enhancement is actually caused by the quasar, it corresponds to the other $1\,\sigma$ contour at large quasar ages ($\tTO > 30\Myr$) and moderate obscured sky fractions.
The single bin statistic indicates $\Oobsc < 60\%$ while the 3-bin statistic prefers $\Oobsc\approx 50\%$. We consider both statistics to be consistent here, given that slightly different parts of the spectra are used for the measurements and the overall high stochasticity of the \ion{He}{ii} transverse proximity effect.    

To summarize, our analysis delivers very different results for the six quasars.
The three quasars with the highest estimated \ion{He}{ii} photoionization rate at the background sightline (HE2QS\,J2149$-$0859, SDSS\,J1101$+$1053, HS\,1157$+$3143) are not associated with \ion{He}{ii} transmission spikes and either young ($\lesssim10\Myr$) or highly obscured ($\Oobsc\gtrsim70\%$). The constraints for the foreground quasar along the Q\,0302$-$003 sightline are almost exactly the opposite, with an age most likely $>15\Myr$ and obscuration $<35\%$. For the two other quasars we infer only very weak
constraints. 
One shows moderate age and moderate obscuration, the other bimodal posteriors. These two objects cause the lowest expected \ion{He}{ii} ionization rate at the background sightline in our sample, only $7.4\times10^{-15}\,\mathrm{s}^{-1}$ and $9\times10^{-15}\,\mathrm{s}^{-1}$. This probably represents the limit for deriving constraints on individual foreground quasars.

\section{Discussion}
\label{Sec:Discussion}
 
Given the surprisingly dissimilar appearance of the \ion{He}{ii} spectra for the six quasars with the highest \ion{He}{ii} photoionization rate at the background sightline presented in \citet{Schmidt2017} and modeled in detail here, 
one might have expected substantial differences in the emission of ionizing radiation for these quasars. 
In this study we quantified this by comparing the transmission spectra of each quasar's background sightline
to detailed models of the transverse proximity effect, parametrized by quasar age ($\tTO$) and degree of obscuration ($\Oobsc$).
Nevertheless, it remains challenging to interpret our results in the context of a single model of quasar emission.

In the simplest picture, all quasars are drawn from the same underlying population with a unique set of properties, in our case $\Oobsc$ and $\tQ$.
From previous studies one might have expected a fiducial quasar model with e.g. $\sim 50\%$ obscuration \citep{Simpson2005, Brusa2010, Assef2013, Lusso2013, Buchner2015, Marchesi2016} and a lifetime of $\sim 25\Myr$ \citep{Schmidt2017}.
This fidicual model lies in the center of the posterior distributions in Figure~\ref{Fig:Key_Plot}. 
Although our confidence contours are relatively broad, such that this parameter combinations is never formally ruled out at high confidence, it is
rather intriguing that none of our posterior distributions actually have a peak at this location in parameter space. 
Of course, a substantial amount of variation around a fiducial model has to be expected and in particular the quasar age is, even for a fixed quasar lifetime $\tQ$, a random variable drawn from $0 < \tTO < \tQ$. The weak and degenerate nature of our constraints and the small number of objects makes it challenging to formally compare the probability of different models. However, visual inspection of the posterior distributions in Figure~\ref{Fig:Key_Plot} suggests that quasars tend to live in two different regions of this parameter space
with dissimilar emission properties, 
suggesting one group being very young or highly obscured and the other old and unobscured.

One can speculate that the position at which a quasar lives in this (bimodal) emission parameter space might somehow correlate with other quasar properties. For instance, in receding torus models \citep{Lawrence1991,Simpson1998, Honig2007} the obscured sky fraction depends strongly on luminosity.  Figure~\ref{Fig:Key_Plot} lists numerous quantities for the six foreground quasars. 
However, there are no obvious trends with quasar properties such as absolute magnitude or redshift, or other parameters like  \ion{He}{ii} ionization rate or separation from the background sightline. 
It would be interesting to investigate the dependence on black hole masses or Eddington ratios. 
However, there are so far no observations enabling measurements of these quantities for the analyzed foreground quasars.	
We therefore at present do not have a convincing explanation for the origin of this suggestive bimodality.

\subsection{Generalization to Quasar Population Properties and Constraints of Additional Parameter}
\label{Sec:PopulationProperties}

On the other hand, assuming that all quasars do represent a common population, it is intriguing to use our analysis of individual quasars to derive properties of the general quasar population. However, this is, given our non-localized posterior probabilities and the strong degeneracies a rather difficult task.
We point out that a proper Bayesian answer to this requires more than just multiplying our individual likelihoods and marginalizing them over one of the two parameter. Instead, such an attempt requires a very careful analysis to avoid introducing any subtle biases, in particular due to the inevitable strong influence of explicit and implicit priors. 
In particular for a lifetime estimate, we would have to distinguish between general lifetime $\tQ$ and individual quasar ages $\tTO$. 
This probably requires drawing the quasar ages from a distribution between $0 <\tTO < \tQ$, which turns $\tTO$ from a deterministic parameter to a random variable, requiring a large number of additional skewers to properly sample the parameter space.
Such an analysis is beyond the scope of this paper. 

One might also wish to include additional parameters in the analysis like the IGM mean free path for \ion{He}{ii} ionizing photons or the ionizing output of quasars. This is in principle possible, it would however substantially complicate the analysis and require the sampling of a much larger parameter grid. In addition, as illustrated in Figure~\ref{Fig:Model_Averages}, the constraining power of the observations is limited due to the high expected variance. The possible constraints from single sightlines would therefore likely be unsatisfactory.

\subsection{Non-Lightbulb Quasar Lightcurves and Non-Equilibrium Effects}
\label{Sec:Non-Lightbulb}

As stated before, we use a lightbulb model for the quasar lightcurve and assume instantaneous photoionization equilibrium. Both aspects are clearly a simplification. 

As described e.g. in \citet{Khrykin2016}, the characteristic timescale helium requires to react to a change in the photoionization rate and adopt a new ionization equilibrium, the equilibration timescale, is rather long. It depends on the recombination and photoionization timescale as $t_\mathrm{eq} = 
{(\,{t_\mathrm{phot}}^{-1} + {t_\mathrm{recom}}^{-1}\,)}^{-1}$.
The recombination timescale $t_\mathrm{recom}^\mathrm{HeII} = {( \alpha_\mathrm{A}^\mathrm{HeII} \, n_{e^-})}^{-1}$ in the IGM of our simulations lies between $1.1\Gyr$ and $3.5\Gyr$.
The \ion{He}{ii} equilibration timescale is therefore always dominated by the photoionization timescale ${t_\mathrm{phot}^\mathrm{HeII}} = {\Gamma_\mathrm{HeII}}^{-1}$, which itself depends on the intensity of the quasars radiation and the UV background.
If our quasars illuminate the background sightline, the photoionization timescale is rather short, about $2.5\Myr$. 
If the background sightline is not illuminated, the UV background determines the photoionization timescale and common values are between  $4.8\times10^{-15}\,\mathrm{s}^{-1}$ and $5.7\times10^{-16}\,\mathrm{s}^{-1}$, corresponding to $6.6\Myr$ and $55\Myr$, respectively (see Figure~\ref{Fig:FredUVB}). 
The latter case might be important if a quasar turns off. It determines how fast a possible transverse proximity effect vanishes.

Neglecting these non-equilibrium effects can have different effects on our measurement.
For three of the four strongest foreground quasars we find no evidence for an influence on the background sightline and explain this with either high obscuration or young quasars ($\tTO<10\Myr$). In the latter case, heuristically
one should add the photoionization timescale of $\approx2.5\Myr$  to this constraint. In the case of the Q\,0302$-$003 $z=3.05$ object, our most-likely quasar age of $22\Myr$ should be longer by a similar amount. 

In cases like Q\,0302$-$003, where we find a clear transverse proximity effect, the exact constraint is only that the quasar is observed today to be  active (as seen on the direct view, $\Delta{}t=0$) and had to be active approximately one transverse light crossing time earlier ($3\Myr \lesssim \Delta{}t \lesssim 30\Myr$, see Figure~\ref{Fig:TimeRetardation}) as probed by the background sightline.
Technically, the quasar could have been inactive in between these times.
If one allows such a quasar lightcurve composed of (at least) two (shorter) bursts, our measurement constrains their separation in time instead of the duration of one long, continuous burst. In addition, the first burst, responsible for the enhanced \ion{He}{ii} transmission at the background sightline, must have been longer than the \ion{He}{ii} equilibration timescale to actually have an impact on the ionization state, thus $>2.5\Myr$. 
Clearly, allowing rather flexible quasar lightcurves and considering non-equilibrium effects, makes an already rich problem even more complicated. 

Another scenario one might consider is \textit{quasar flickering} \citep[e.g.][]{Novak2011, Segers2017}. In such a case, the quasar switches rapidly, e.g. with periods of $10^5\yr$, between on and off states.
For the \ion{He}{ii} ionization state only the ionizing flux averaged over the equilibration timescale is
relevant. Flickering on timescales $10^5\,\mathrm{yr}$, much shorter than $t_\mathrm{eq}$, would therefore be indistinguishable from continuous emission with the quasar luminosity reduced by the appropriate amount.
However, flickering with the cosmic average duty cycle of $\approx1\%$ \citep{Conroy2013, Eftekharzadeh2015} would probably not provide sufficient ionizing photons to cause an observable effect.
In \citet{Schmidt2017} we found statistical evidence for a transverse proximity effect for estimated photoionization rates $\GqsoHe > 2\times10^{-15}\,\mathrm{s}^{-1}$, roughly comparable with the UV background. If the actual, time-averaged \ion{He}{ii} ionizing flux output from quasars were lower by more than a factor of a few, the quasars would not cause any significant enhancement over the UV background and no proximity effect would be visible. This sets limits to the minimum duty cycle of a possible quasar flickering.  The cosmic average of 1\% would certainly be too low. The quasars therefore would have to be, despite their flickering, in an extended phase of high activity and our measurement constrains the duration of this phase.

For arbitrary or very complicated lightcurves \citep[e.g.][]{Novak2011} it becomes challenging to arrive at firm constraints. With a proper parametrization of the quasar lightcurve, radiative transfer calculations could in principle deliver the required models, but given the large amount of expected scatter in \ion{He}{ii} transverse proximity effect measurements (Figure~\ref{Fig:Model_Averages}), it appears unlikely to derive meaningful results. Instead of developing more sophisticated models, it seems more appropriate to focus future efforts towards reducing the variance in the measurement.

\section{Summary}

In \citet{Schmidt2017} we presented the results of our dedicated \ion{He}{ii} foreground quasar survey and provided statistical evidence for the presence of the \ion{He}{ii} transverse proximity effect, which resulted in a heuristic constraint on the quasar lifetime $\tQ > 25\Myr$.
However, among the six foreground quasars with the highest \ion{He}{ii} photoionization rates, only one is associated with a strong \ion{He}{ii} transmission spike. For the other three, no comparable signature on the background sightline is observed, which might point towards very young or highly obscured quasars.

In this study we investigate the implications of these high photoionization rate sightlines via detailed modeling of the \ion{He}{ii} transverse proximity effect, encompassing finite quasar ages $\tTO$,  light travel time effects, opening angle/obscuration $\Oobsc$, and stochasticity of both the IGM and quasar orientation.
We use outputs from the \Nyx cosmological hydrodynamical simulations \citep{Almgren2013, Lukic2015} and post-process these with the fluctuating 
\ion{He}{ii} UV background model from \citet{Davies2017} (see Figure~\ref{Fig:FredUVB}) plus the added effect of one isolated foreground quasar. The UV background model is calibrated to match existing \ion{He}{ii} observations (Figure~\ref{Fig:Test_UVB}).  
For the foreground quasar, we vary quasar age $\tTO$ and obscuration $\Oobsc$ to explore their combined effect on the \ion{He}{ii} transverse proximity effect signal, as well as obtain the first estimates of its variance resulting from IGM density fluctuations, UV background fluctuations, and the unknown orientation of the foreground quasars (Figure~\ref{Fig:Model_Averages}).
We adopt a fully Bayesian statistical approach to deal with the large non-Gaussian fluctuations in the expected \ion{He}{ii} transmission, Poisson photon-counting noise, and strong parameter degeneracies (\S~\ref{Sec:4}).

We derive joint constraints on $\tTO$ and $\Oobsc$ (Figure~\ref{Fig:Key_Plot}), for the six \citet{Schmidt2017} foreground quasars with the highest \ion{He}{ii} photoionization rates.  A highly inhomogeneous picture of quasar emission properties emerges from this analysis.  
For the prototype quasar associated with the \ion{He}{ii} transmission spike in the Q\,0302$-$003 sightline, our analysis prefers $\tTO\approx22\Myr$ and low obscuration ($\Oobsc<35\%$). For three other foreground quasars however, we rule out long lifetimes ($>10\Myr$) and low obscuration $\Oobsc<60\%$. Although a fiducial quasar model with $\tTO=25\Myr$ and 50\% obscuration is marginally consistent ($2\sigma$) with most of our derived probability contours, our analysis indicates a bimodal distribution of quasar properties with one group being old and nearly unobscured while the other one is very young or highly obscured. 
An analysis of trends with other quasar parameters, e.g. luminosity, redshift, \ion{He}{ii} ionization rate, does not lead to a convincing explanation for the origin of this apparent dichotomy.

Further progress in deriving constraints on quasar or IGM properties is hindered by the large intrinsic variance of the \ion{He}{ii} transverse proximity effect as shown in Figure~\ref{Fig:Model_Averages}.
This poses a fundamental limitation for similar studies of the \ion{He}{ii} transverse proximity effect.
A possible solution to this issue could be the statistical combination of individual measurements as discuss in \S~\ref{Sec:PopulationProperties} or stacking as in \citet{Schmidt2017}. However, the available foreground quasar sample is limited and including fainter quasars increases sensitivity to the exact details of the fluctuating \ion{He}{ii} UV background model.
Alternatively, the discovery of individual foreground quasars with substantially higher \ion{He}{ii} photoionization rate than the quasars analyzed in this study might offer a viable opportunity to derive firmer constraints. Here, the transverse proximity effect would be stronger, reducing the relative uncertainty.
However, despite our survey efforts in \citet{Schmidt2017}, such objects could so far not be discovered.
Maybe the best option to overcome the intrinsic IGM variance associated with the \ion{He}{ii} transverse proximity effect could be the use of coeval hydrogen $\lya$ forest spectra.
In principle, high-resolution \ion{H}{i} absorption spectra could deliver information about the local IGM density structure and possibly allow a more precise measurement of the \ion{He}{ii} ionization state. 
Developing models and a statistical framework to exploit this additional information constitutes an interesting task for the future.

\section*{Acknowledgments}

We would like to thank the members of the ENIGMA\footnote{\url{http://enigma.physics.ucsb.edu/}} group at the Max Planck Institute for Astronomy (MPIA) and UCSB for useful discussions and support. GW has been supported by the Deutsches Zentrum f\"ur Luft- und Raumfahrt (DLR) under contracts 50\,OR\,1317 and 50\,OR\,1512.

\bibliographystyle{APJ}
\bibliography{Literature}

\section*{Appendix}

\subsection*{Likelihood Calculation for $\xiTPE$ Statistic}

Calculating the likelihood using $\xiTPE$ is in large parts identical to the approach outlined above with just replacing $\F$ by $\xiTPE$. However, Equations~\ref{Eq:3} to \ref{Eq:4} describing the measurement process on the observed spectra have to be modified to include the uncertainty in the background estimate. The first term in Equation~\ref{Eq:2} can be written as

\begin{equation}
p( \CT | \xiTPE ) = \prod_i \int p( \CT_i | \xiTPE, \F_\mathrm{BG} )  \; p( \F_{BG} ) \; d{\F_\mathrm{BG}}
\label{Eq:7}
\end{equation}

where $\F_\mathrm{BG}$ denotes the intrinsic transmission in the region $15\cMpc<|R_\parallel|<65\cMpc$ and $p( \CT | \xiTPE, \F_\mathrm{BG} )$ is again the Poisson probability for the photon counts in the proximity region
\begin{equation}
P_{\CML_i}(\CT_i) = \frac{ {\CML_i}^{\,\CT_i} }{\CT_i !} \: \mathrm{e}^{-\CML_i}
\label{Eq:8}
\end{equation}
with $\CML_i$ now expressed in terms of $\xiTPE + \F_\mathrm{BG}$ \\
\begin{equation}
\CML_i = (\xiTPE + \F_\mathrm{BG}) \cdot \CO_i \cdot \SE_i \cdot \EX_i + \BG_i 
\label{Eq:9}.
\end{equation}

Although the background transmission is calculated over $3\times$ the pathlength of the the $\pm15\cMpc$ proximity region and the photon-counting noise therefore far less important, we fully propagate the associated uncertainty. In practice, the probability for the background transmission from Equation~\ref{Eq:7} can be computed in the following way, using Bayes' theorem:  
\begin{equation}
p( \F_\mathrm{BG} | \CT^\mathrm{BG} ) \propto \prod_j p( \CT_j | \F_\mathrm{BG} ) \; p_0( \F_\mathrm{BG} )
\label{Eq:10}.
\end{equation}
The first term is again the Poisson probability for measuring counts $\CT^\mathrm{BG}=\{\CT_j\}$ now in the background region given $\F_\mathrm{BG}$ and the second term is a prior for the background transmission $p_0( \F_\mathrm{BG} )$ we had to introduce. We adopt a flat prior but the exact choice is unimportant. The background transmission is estimated over a large pathlength and fairly well constrained from the data.

The remaining part of the likelihood calculation is completely analogous and just requires to measure $\xiTPE$ in the simulated skewers instead of $\F$.

\end{document}